\documentclass [12pt]{article}
\usepackage{lscape}                                           %
\usepackage[centertags]{amsmath}
\usepackage{amsfonts} \usepackage{amssymb}\usepackage{eufrak}
\usepackage{graphicx}
\usepackage{a4}
\usepackage{cite}
\usepackage{lscape}

\begin{document}
\begin{titlepage}
\rightline{}
\rightline{}
\vskip 3.0cm
\centerline{\LARGE \bf Comparing Double String Theory Actions}

\vskip 1.0cm \centerline{\bf L. De Angelis$^a$, G. Gionti S. J.$^b$,  R. Marotta$^c$ and F. Pezzella$^c$}
\vskip .6cm
\vskip .4cm \centerline{\sl $^a$ Dipartimento di Fisica, Universit\`a degli Studi ``Federico II'' di Napoli}
\centerline{\sl Complesso Universitario Monte S. Angelo ed. 6, via Cintia,  80126 Napoli, Italy}
\vskip .4cm
\centerline{\sl
 $^b$ Specola Vaticana}
\centerline{\sl Vatican City, V-00120, Vatican City State and Vatican Observatory Research Group}
\centerline{\sl Steward Observatory, The University Of Arizona, 933 North Cherry Avenue}
\centerline{\sl Tucson, Arizona 85721, USA}
\vskip .4cm
\centerline{\sl $^c$ Istituto Nazionale di Fisica Nucleare, Sezione di Napoli}
\centerline{ \sl Complesso Universitario di Monte
S. Angelo ed. 6, via Cintia,  80126 Napoli, Italy}
\vskip 0.4cm

\begin{abstract}
Aimed to a deeper comprehension of a manifestly T-dual invariant formulation of string theory, in this paper a detailed comparison between the non-covariant action proposed by Tseytlin and the covariant one proposed by Hull is done. These are obtained by  making  both the string coordinates and their duals explicitly appear, on the same footing,  in the world-sheet action, so ``doubling''  the string coordinates along the compact dimensions.  After a discussion on the nature of the constraints in both the models and the relative quantization, it results that the string coordinates and their duals  behave  like ``non-commuting'' phase space coordinates but their expressions in terms of Fourier modes generate the  oscillator algebra of the standard bosonic string.  A proof of the equivalence of the two formulations is given.  Furthermore, open-string solutions are also discussed.

\end{abstract}

\end{titlepage}

\newpage

\tableofcontents
\vskip 1cm

\section{Introduction}
It is well-known that in order to connect string theory to real-world physics, it has to be compactified from ten (twentysix, if only the bosonic theory is considered) to four dimensions. For closed strings, the presence of $D$ compact dimensions $X^{a}$  implies the existence not only of momentum modes $p_{a}$ which are quantized along such dimensions,  but also of winding modes $w^{a}$ representing the number of times the string winds around the compact dimension. Topologically, the closed string winding number is a meaningful concept.

Just as $p_{a}$ can be considered as the momentum associated with $X^{a}$, one can ask what is the coordinate the winding number $w^{a}$ is associated with. The answer to this question is provided by $\tilde{X}_{a}$, the T-dual coordinate of $X^{a}$, which is a co-vector (one-form) being  $w^{a}$ a vector.

T-duality is an old subject in string theory (for a recent review, see Ref. \cite{1302.1719} and references therein). It implies that in many cases two different geometries for the extra dimensions are physically equivalent.
T-duality is therefore a clear indication that ordinary geometric concepts can break down in string theory at the string scale. In the simplest case of a circle compactification, it implies that the closed string compactified on a circle of radius $R$ is equivalent to the one compactified on a  circle of radius $\alpha'/R$. But more than a mere duality, T-duality is an exact symmetry of the Hamiltonian, and hence of the spectrum, of a closed string compactified on a circle. In this case, T-duality is encoded in the simultaneous transformations $R \leftrightarrow \alpha'/R$ and $p_{a}  \leftrightarrow w^{a} / \sqrt{\alpha'}$  under which $X^{a} \leftrightarrow \tilde{X}_{a}$, with $w^{a}$ playing the role of momentum mode for $\tilde{X}_{a}$. The fact that T-duality is an exact symmetry for closed strings suggests that one could extend the standard formulation,  based on the Polyakov action, by  introducing the symmetry at the level of the world-sheet sigma-model Lagrangian density, so looking for a manifestly T-dual invariant formulation of closed string theory. This, of course, requires the introduction, in the sigma-model,  of {\em both} the compact coordinates $X^{a}$ {\em and} the dual ones $\tilde{X}_{a}$, so it is based on a {\em doubling} of the string coordinates in the target space, hence the name of {\em double string theory}. It appears that the compact part of the target space in {\it double string theory} is locally defined by the direct sum of the tangent and cotangent spaces in each point.


The main goal of this new action would be to explore more closely the gravity implied by string theory. In fact, if interested in writing down the complete effective field theory of such generalized sigma-model, one should consider, correspondently to the introduction of $X^{a}$ and $\tilde{X}_{a}$, a dependence of the fields associated with string states on such coordinates, besides the one on the non-compact dimensions. So one can claim that the {\em double string effective field theory} is  a {\em double field theory} \cite{0904.4664, 1003.5027, 1309.2977, 1306.2643,1207.6110,0708.2267, 0712.1121, 1305.1907}. In particular, this has to be true for the well-known effective gravitational action of a closed string involving the fields associated with its massless states: the gravitational field $G_{\mu \nu}$, the Kalb-Ramond field $B_{\mu \nu}$ and the dilaton $\phi$. So one can ask what this action becomes in light of the fact that all those fields depend on $X^{a}$ and $\tilde{X}_{a}$ and, in particular, which symmetries and what properties it would have, perhaps shedding light on aspects of string gravity unexplored thus far. But, of course, in order to answer these questions, one must first find an answer to the more fundamental question of how the closed string would look like when the T-duality is manifested in the sigma-model Lagrangian density.

First attempts to face these issues were already explored by W. Siegel in Ref. \cite{WS} and by A. A. Tseytlin in Refs. \cite{TPL, TNP}. In particular, the latter author defines a sigma-model action written in a first-order form involving string coordinates mapping the string in the compact factor ${\cal M}$ of the target space ${\mathbb R}^{1,d-1} \otimes {\cal M}$, besides the usual string coordinates mapping the string in the uncompact Minkowski factor  ${\mathbb R}^{1,d-1}$. This model is essentially described by the sum of actions for the right and left scalar string coordinates $X_{R;L}$  reproducing the Floreanini-Jackiw Lagrangians respectively for antichiral and chiral scalar fields. It is not manifestly local Lorentz invariant, but this invariance is recovered on-shell.  In fact, it is precisely the requirement that the local Lorentz invariance could hold on-shell to dictate a constraint in this model that implies the geometry of the double torus determined by the $O(D,D)$ invariant metric. This invariance results to be, therefore, an {\em output} of the theory coming from its consistency. As a result of this symmetry, the non-covariant action contains the $O(D,D)$ invariant metric together with a generalized target space metric depending on $D^{2}$ moduli which are identified with the background values of the components of the fields $G$ and $B$.

In this paper a review of this approach is first given.  Then the Dirac method of quantizing constrained systems is applied to this theory, since it contains primary second class constraints.

The Dirac procedure is carried out in the convenient basis provided by  the right and left coordinates $X_{R;L}$, where both the $O(D,D)$ and the generalized metrics are diagonal. In such basis, all of the explicit dependence on the $B$-field disappears,
making the analysis easier, but it can be reintroduced by  any $O(D,D)$ rotation.
The presence in the theory of second class constraints leads to the introduction of the Dirac brackets and the quantization is performed by substituting the latter with commutators, as usual. It turns out that the mode expansions of the fields $X_{R;L}$ satisfy the same commutation relations as the ones of the string modes.  Then, Virasoro generators are introduced: they provide constraints coming from the equations of motion of the zweibein.
This procedure will lead to the interesting result that the coordinates $X^{a}$ and $\tilde{X}_{a}$ behave  like non-commuting phase space coordinates \cite{1010.1361,1211.6437} but their expressions in terms of Fourier modes generate the usual oscillator algebra of the standard formulation.

Besides the non-covariant double string theory \`a la Tseytlin, a covariant version has been proposed by C. Hull \cite{0406102} in which the $O(D,D)$ invariance is an {\em input} of the theory. More precisely, the author starts with a covariant action already involving a doubled number of string coordinates on the torus, exhibiting the manifest  $GL(2D;{\mathbb Z})$ invariance that, in turn, generates the $O(D,D)$ symmetry when a self-duality constraint is imposed, halving the degrees of freedom.

 In this paper, a comparison between the two approaches will be carefully done and, in particular, it will be shown that the constraint imposed by Hull is equivalent to the one of Tseytlin for restoring the local Lorentz invariance. Furthermore, it will be explicitly shown that introducing the Hull's constraint in the covariant action, according to the procedure introduced by Pasti, Sorokin and Tonin \cite{PST1, PST2} reproduces the non-covariant action (see also Refs. \cite{0708.2267, 1308.4418}).   The connection between the two formulations has already been noticed in Refs.\cite{1306.2643, 1308.4418} in the case of one compact dimension and in the absence of the $B$-field. It is here generalized for $D$ compact dimensions and in the presence of a non-trivial background.
This result clearly shows that the two models are equivalent.
Also for the covariant action, a careful analysis of the quantization,  initiated  in Refs.  \cite{0605114, 0701080, 1111.1828, 1106.1888, 1306.2970}, is performed.
Here, it  is carried out in the $X_{R;L}$-frame  where the Dirac quantization  can be straightforwardly made in the general case.  The duality constraints satisfy the same algebra as the primary second class constraints of the non-covariant model. Hence, Dirac brackets are introduced: these,  once  replaced by commutators, lead for the Fourier modes of the fields $X_{R;L}$ to the same commutation relations as the ones in the Tseytlin model.  Finally, it is shown that the quantization of the Hull covariant model is exactly the same as the Tseytlin non-covariant model.

Manifestly T-duality invariant models were originally proposed in the framework of closed string theories. However, suggestions on how to include open strings with D-branes \cite{0406102, 1306.2643} and superstrings have also been proposed \cite{1106.4015, 1308.4829}. 
In the same spirit, it has been explored the possibility of canceling out the surface integrals generated from the derivation of the equations of motion, by imposing open-string like boundary conditions. These relate $X^{a}$ and $\tilde{X}_{a}$ on the world-sheet boundaries.  The analysis has been done in the basis of the right and left coordinates and the boundary conditions imposed on these quantities result to be the same as the ones usually imposed on the corresponding bosonic string fields in the presence of a magnetic field\cite{0709.4149}.

The structure of this paper is the following.

Sect. 2 is devoted to the non-covariant double string sigma-model first introduced by Tseytlin. In particular, in subsect. 2.1 the action and its symmetries will be described. Explicit solutions of the equations of motion for the string coordinates are given. In subsect. 2.2 the analysis of the constraints will be performed in the presence of second-class constraints leading to Dirac brackets. After that, quantization is discussed.

Sect. 3 is devoted to the covariant double string sigma-model introduced by Hull. The relative action, its symmetries and its constraints will be analyzed and a demonstration of its equivalence with the non-covariant action is done. After the analysis of such constrained system, its quantization will be faced.

In Sect. 4,  explicit open string solutions of the equations of motion for the string coordinates are given, together with  a more intuitive picture of what ``dual field'' could mean in this case.

Three Appendices complete this work. In Appendix A, notations are fixed and useful identities used in the text are summarized. In Appendix B, details on solving the equations of motion in both the approaches are given, together with some details on the quantization procedure. In Appendix C, the open string symmetry $O(D)$ is examined.

\section{The non-covariant double string sigma-model}   \label{tseytlin}

\subsection{Action and its symmetries}

The aim of this section is to review the non-covariant T-duality symmetric formulation \cite{TPL, TNP} of the bosonic string theory. 

The starting point is the following generalized sigma-model action:
\begin{eqnarray}
S[e^a_{~\alpha}, \chi^i]= - \frac{1}{2} \int_{\Sigma} d^2 \xi \,\,  e\,\,  {\cal C}_{i j}^{a b}(\chi) \nabla_a \chi^i \nabla_b \chi^j  \label{act0}
\end{eqnarray}
where the coordinates on the two-dimensional manifold $\Sigma$ are $ \xi^0 \equiv \tau, \xi^1 \equiv \sigma$.  It is a functional of the zweibein $e_{\,\, \alpha}^{a}(\xi)$, being $a$  and  $\alpha$, respectively, the label for the flat  and the curved index,  and of  $N$ two-dimensional scalar fields $\chi^i (\xi)$  which are vectors in an $N$-dimensional target space ${\cal M}$. Furthermore,  $\nabla_{a} \chi^{i} = e_{a}^{ \,\, \alpha} \partial_{\alpha} \chi^{i}$ and $e=\mbox {det} \, [ e_{\,\, \alpha}^{a}]$.

The action (\ref{act0}) is meant to be generic, with the number of embedding coordinates $\chi^{i}$ kept, at this level, unspecified.  Indeed, the usual sigma-model action for strings propagating in a background is obtained considering ${\cal C}_{i j}^{a b} = T (\eta^{ab}G_{ij} - \epsilon^{ab} B_{ij})$ $(\epsilon^{01}=-\epsilon^{10}=1)$, being $T$ the string tension,  $G_{ij}$ the metric tensor of the target space and $B_{ij}$ the antisymmetric Kalb-Ramond field. In this case the scalar fields $\chi^{i}$  $(i=1, \dots, N)$ are the string coordinates in ${\cal M}$. The same action will be suitable, under certain conditions,  to describe a "double
 string" sigma-model with manifest T-duality, as we are going to show.

Let us consider the case in which the action (\ref{act0})  can be  rewritten in a first order form \cite{TNP}  independently of the value taken by the coefficients ${\cal C}_{ij}^{00}$ that will be considered vanishing since now on. One gets:
\begin{eqnarray}
S=-\frac{1}{2} \int d^2 \xi \,\,  e\,\, \left[ \mbox{\bf{C}}_{ij} \nabla_{0} \chi^{i} \nabla_{1} \chi^{j} + M_{ij} \nabla_{1} \chi^{i} \nabla_{1} \chi^{j} \right] \, ,
\end{eqnarray}
with $\mbox{\bf{C}}_{ij} =  {\cal C}_{i j}^{0 1} +  {\cal C}_{ji}^{1 0}$ and $M_{i j} =M_{ji}\equiv {\cal C}_{ij}^{11}$.

Rewriting $\mbox{\bf{C}}_{ij} = \mbox{\bf{C}}_{(ij)} +\mbox{\bf{C}}_{[ij]} \equiv C_{ij} + H_{ij}$ yields to:
\begin{eqnarray}
S= - \frac{1}{2} \int d^{2} \xi \, e \left[ C_{i j} \nabla_{0} \chi^{i} \, \nabla_{1} \chi^{j} + \frac{1}{2} \epsilon^{a b} H_{ij} \nabla_{a} \chi^{i} \nabla_{b} \chi^{j} + M_{i j} \nabla_{1} \chi^{i} \nabla_{1} \chi^{j} \right] .   \label{act1}
\end{eqnarray}
The action (\ref{act1}) exhibits the following local invariances:
\begin{itemize}
\item invariance under two-dimensional diffeomorphisms $\xi^{\alpha} \rightarrow \xi'^{\alpha}( \xi)$ acting as
\begin{eqnarray}
\chi'^{i} (\xi'^{\alpha}) = \chi^{i} (\xi^{\alpha})
\,\,\,\,\, \mbox{and} \,\,\,\,\,
e'^{a}_{\,\,\, \alpha} = e^{a}_{\,\,\, \beta} \frac{\partial \xi^{\beta}}{\partial \xi'^{\alpha}} \,\, ;
\end{eqnarray}

\item
invariance under Weyl transformations
\begin{eqnarray}
e^{a}_{\,\, \alpha} \rightarrow \lambda (\xi) e^{a}_{\,\, \alpha}  \, ,
\end{eqnarray}
which leave the fields $\chi^{i}$ and the quantities $e\, e_{a}^{\, \alpha} \, e_{b}^{\, \beta}$ invariant.
\end{itemize}

Generally, when a vielbein is introduced, then one must ensure that the formalism is invariant under local Lorentz transformations, so that physical observables are independent of the arbitrary choice of the vielbein itself. In fact, as good as $e^{a}_{\, \alpha}$ would be
\begin{eqnarray}
e'^{a}_{\,\,\, \alpha} = \Lambda^{a}_{\,\,b} (\xi) e^{b}_{\,\, \alpha} \, ,
\end{eqnarray}
with $\Lambda^{a}_{\,b} (\xi) $ being an arbitrary $\xi$-dependent Lorentz $SO(1,1)$ matrix. This finite transformation on $e^{a}_{\, \alpha}$  induces the following infinitesimal one:
\begin{eqnarray}
\delta e^a_{~\alpha}=  \omega^{a}_{~b} (\xi)  e^{b}_{~\alpha} \, ,
\end{eqnarray}
with $\omega_{ab} = - \omega_{ba}$. In particular, the choice $\omega^a_{\,\, b}(\xi) = \alpha(\xi)\epsilon^{a}_{~b}$ will be here performed. The action (\ref{act1}) is not manifestly invariant under such transformations, so the requirement of {\em on-shell} local Lorentz invariance has to be made.

In order to study the variation of the action under local Lorentz transformations one can neglect, in fact, the only term having such a symmetry, that is the one proportional to $H_{ij}$. This simplifies the action as follows:
\begin{eqnarray}
S= - \frac{1}{2} \int d^{2} \xi \, e \left[ C_{ij} \nabla_{0} \chi^{i} \, \nabla_{1} \chi^{j}  + M_{i j} \nabla_{1} \chi^{i} \nabla_{1} \chi^{j} \right].  \label{act}
\end{eqnarray}
It results that the variation of $S$ under an infinitesimal local Lorentz transformation $\delta e^{a}_{~ \alpha} = \alpha (\xi) \epsilon^{a}_{~ b} e^{b}_{~ \alpha}$ is
\begin{eqnarray}
\frac{ \delta S}{\delta e^{a}_{\,\, \alpha} } \delta e^{a}_{\,\, \alpha} = \alpha(\xi) \frac{ \delta S}{\delta e^{a}_{\,\, \alpha} } \epsilon^{a}_{\,\, b} e^{b}_{\,\, \alpha} \label{varia}
\end{eqnarray}
and can be expressed in terms of the $\epsilon$-trace ($\hat{t} \equiv \epsilon^{a}_{\,\,b}\, t_{a}^{\,\,b}$) of  the tensor $t_{a}^{\,\,b}$ so defined:
\begin{eqnarray}
t_{a}^{~b} \equiv  \frac{2}{e}  \frac{\delta S}{\delta e_{~\alpha}^{ a}} e_{~\alpha}^{b} .
\end{eqnarray}
The explicit expression for $t_{a}^{\,\,b}$ can be straightforwardly computed from the action (\ref{act}) and it results to be:
\begin{eqnarray}
t_{a}^{\,\,b} & = & - \delta^{b}_{a} \left[ C_{ij} \nabla_{0} \chi^{i} \nabla_{1} \chi^{j} + M_{ij} \nabla_{1} \chi^{i} \nabla_{1} \chi^{j} \right] \nonumber \\
&&  +\delta_{0}^{b} C_{ij} \nabla_{a} \chi^{i} \nabla_{1} \chi^{j} + \delta_{1}^{b} C_{ij} \nabla_{0} \chi^{i} \nabla_{a} \chi^{j} + 2 \delta_{1}^{b} M_{ij} \nabla_{a} \chi^{i} \nabla_{1} \chi^{j} . \label{tab}
\end{eqnarray}
The vanishing of the variation (\ref{varia}) is equivalent to the condition
\begin{eqnarray}
\epsilon^{ab} t_{a b}=0 \, .
\end{eqnarray}
Furthermore, the Weyl invariance implies:
\begin{eqnarray}
t_a^{~a} =\mbox{Trace}\,\, [t_{a}^{\,\, b}] = 0
\end{eqnarray}
since:
\begin{eqnarray}
0= \frac{\delta S}{\delta e_{~\alpha}^{a} } \lambda  e_{~\alpha}^{a} =  \frac{\lambda}{2} e  t_{a}^{\,\, a} \,\, .
\end{eqnarray}
One can easily see from eq. (\ref{tab}) that $t_{00}=t_{11}$, as it must be since the theory is Weyl invariant.

The equation of motion for $e^{a}_{~\alpha}$, $ \delta S / \delta e_{~\alpha}^{a}=0$, implies
\begin{eqnarray}
 t^{~b}_{a}=0 \, .  \label{eqmot}
\end{eqnarray}
This is similar to what happens in the usual formulation of string theory, where the equation of motion for the world-sheet metric $g_{\alpha \beta}$ ($\delta S / \delta g_{\alpha \beta}=0$) determines the vanishing of the energy-momentum tensor $T_{\alpha \beta} \equiv - \frac{2}{T} \frac{1}{\sqrt{-g}}  \frac{\delta S}{\delta g^{\alpha \beta}} $. Eq. (\ref{eqmot}) has to be imposed as an additional constraint both at the classical and at the quantum level.

As previously shown, the requirement of local Lorentz invariance implies the  vanishing of the $\epsilon$-trace of $t_{ab}$. Hence, on the solution of the equation of motion of the zweibein (\ref{eqmot}), this condition is satisfied and the local Lorentz invariance is recovered. The invariances under diffeomorphisms and Weyl transformations, together with this latter invariance that holds on-shell, allow to choose the {\em flat gauge} $e^{a}_{~\alpha} = \delta^{a}_{\alpha}$ for the zweibein. The analogy with the usual formulation of string theory is very strong. In that case the equation of motion for the world-sheet metric, $T_{\alpha \beta}=0$, play the role of constraints while the conformal gauge in which $g_{\alpha \beta} = \eta_{\alpha \beta}$ plays the same role as the flat gauge.

The equation of motion for $\chi^{i}$ is now going to be considered in the case in which the matrices $C$ and $M$ are constant. Details on the derivation of such equation  are given in Appendix B. Here only the result is quoted:
\begin{eqnarray}
\partial_\alpha  \left[ e_{1}^{~\alpha} e (C_{ij} \nabla_{0} \chi^{j} + M_{ij} \nabla_{1} \chi^{j}) \right]  = 0 \label{1}
\end{eqnarray}
with the following surface integrals:
\begin{eqnarray}
-\left. \int_{-\infty}^{+\infty} d\tau  \delta \chi^i \,e\, e_{1}^{~1} \left( C_{ij}\nabla_0\chi^j+M_{ij}\nabla_1 \chi^j \right)  \right|^{\sigma=\pi}_{\sigma=0} +\frac{1}{2} \left.  \int_{- \infty}^{+ \infty} d \tau \,  C_{ij} \,   \partial_{0} \chi^{j} \delta \chi^{i}  \right|^{\sigma=\pi}_{\sigma=0} \,\, .
\end{eqnarray}
It is crucial, at this point, to observe that, with $C$ and $M$ constant, the action (\ref{act}) has a further local gauge symmetry under the following transformations:
\begin{eqnarray}
\chi^{i} \rightarrow \chi'^{i}= \chi^{i} + f^{i} (\tau, \sigma) \, , \label{fursym}
\end{eqnarray}
with the functions $f^{i}$ satisfying  $ \nabla_{1} f^{i} = 0$ and the same boundary conditions as the fields $\chi$ and $\chi'$. This shift symmetry leaves the equation of motion  in (\ref{1}) invariant. In fact, it generates a vanishing extra term:
\begin{eqnarray}
\partial_{\alpha} \left[ e\,C_{ij} e_{0}^{~\beta}\, \partial_{\beta} f^{j} \right] = \partial_{\alpha} \left[ e\, \nabla_{1} f^{j} C_{ij} \right]  = 0
\end{eqnarray}
where the identity
\begin{eqnarray}
e^{\,\, \alpha}_{0} e^{\beta}_{1} - e^{\,\, \alpha}_{1} e^{\,\, \beta}_{0} = \frac{1}{e} \epsilon^{\alpha \beta} \,\,.\label{iddet}
\end{eqnarray}
has been used. In Appendix B it is shown that the Lagrangian density is modified by a total derivative against the transformation (\ref{fursym}). This symmetry constitutes a relevant aspect of the action (\ref{act}) since it will provide a gauge choice in which the equation of motion becomes of first-order.

In the flat gauge, eq. (\ref{1}) reduces to:
\begin{eqnarray}
\partial_{1} \left[ C_{ij} \partial_{0} \chi^{j} + M_{ij} \partial_{1} \chi^{j} \right] = 0
\end{eqnarray}
from which one obtains:
\begin{eqnarray}
C_{ij} \partial_{0} \chi^{j} + M_{ij} \partial_{1} \chi^{j} = g_{i} (\tau) \, , \label{fi}
\end{eqnarray}
being $g_{i} (\tau) $ an arbitrary $\tau$-dependent function. In particular, the shift symmetry can 
be here used to fix $C\,\partial_0f=g$. As a result  one has:
\begin{eqnarray}
C_{ij} \partial_{0} \chi^{j} + M_{ij} \partial_{1} \chi^{j} = 0 \,\,  \label{fi0}
\end{eqnarray}
and the boundary conditions, once the latter equation is used, reduce to:
\begin{eqnarray}
\frac{1}{2} \left.  \int_{- \infty}^{+ \infty} d \tau \,  C_{ij} \,  \left[ \partial_{0} \chi^{j} \delta \chi^{i} \right] \right|^{\sigma=\pi}_{\sigma=0} \, .
\end{eqnarray}
This term is vanishing when periodicity in $\sigma$ is imposed on $\chi^{i}$ (as it happens for closed strings) or, alternatively, when $ \partial_{0} \chi^{i}=0$ at $\sigma=0, \pi$ (as it happens  for open strings with Dirichlet conditions).

Eq. (\ref{fi0}) in fact appears in the explicit expression of the $\epsilon$-trace of $t_{ab}$. Indeed, computing the $\epsilon$-trace and imposing its vanishing yield to:
\begin{eqnarray}
\epsilon^{ab} t_{a b}  &= & \left[ \nabla_{0} \chi^{i} C_{ij} + \nabla_{1} \chi^{i} M_{ij} \right](C^{-1})^{jk} \left[C_{kl}\nabla_{0} \chi^{l}+M_{kl} \nabla_{1} \chi^{l} \right] \nonumber \\
&& + \, \,  \nabla_{1} \chi^{i} \,\, (C-MC^{-1}M)_{ij}  \nabla_{1} \chi^{j} = 0 \, .  \label{epstrace}
\end{eqnarray}
Hence, in the flat gauge and along the solutions of the equations of motion for $\chi^{i}$, eq. (\ref{epstrace}) reduces to the following condition on the matrices $C$ and $M$:
\begin{eqnarray}
C=MC^{-1}M .  \label{2}
\end{eqnarray}

The matrix $C$ can be always put, after suitably rotating and rescaling $\chi^{i}$, in the following diagonal form:
\begin{eqnarray}
C= \mbox{diag} (1, \cdots, 1, -1, \cdots, -1) \, , \label{24}
\end{eqnarray}
with $p$ eigenvalues $ 1$ and $q$ eigenvalues $ -1$. Being $C=C^{-1}$, this implies that the property in eq. (\ref{2}) becomes the one defining the indefinite orthogonal group $O(p,q)$ of $N \times N$ matrices $M$  with $N=p+q$ (with $p, q$ still undetermined at this level) in $R^{p,q}$ with the standard inner product given by:
\begin{eqnarray}
C=MCM .  \label{2bis}
\end{eqnarray}
With this identification of $C$ and with $\chi^{i} =(\chi_{-}^{\mu},  \chi_{+}^{\nu})$, the action (\ref{act}) can be rewritten as follows:
\begin{eqnarray}
S= - \frac{1}{2} \int d^{2} \xi \, e \left[ \sum_{\mu=1}^{p} \nabla_{0} \chi^{\mu}_{-} \, \nabla_{1} \chi^{\mu}_{-} - \sum_{\nu=1}^{q} \nabla_{0} \chi^{\nu}_{+} \, \nabla_{1} \chi^{\nu}_{+}  + M_{i j} \nabla_{1} \chi^{i} \nabla_{1} \chi^{j} \right]    \label{actchir}
\end{eqnarray}
and it will be shown in a while that it can be interpreted, when a suitable frame is chosen, as describing a system of interacting $p$ two-dimensional antichiral scalar fields ($\dot{\chi}_{-}= - \chi'_{-}$) and $q$ two-dimensional chiral scalar fields ($\dot{\chi}_{+} = \chi'_{+}$), according to the Floreanini-Jackiw Lagrangians for two-dimensional chiral and antichiral scalars \cite{FJ}:
\begin{eqnarray}
{\cal L}_{\pm}( \dot{\chi}_{\pm}\,,\, \chi'_{\pm} ) =  \pm \frac{1}{2} \dot{\chi}_{\pm} \chi'_{\pm} - \frac{1}{2} \chi'^{2}_{\pm}  \label{FJL}.
\end{eqnarray}
Requiring the  absence of a quantum Lorentz anomaly implies that $p=q=D$ with $2D=N$ \cite{AGW, CFR}. Consequently, the matrix $C$ in eq. (\ref{2}) becomes the $O(D,D; {\mathbb R})$ invariant metric in the $2D$-dimensional target space ${\cal M}$ with coordinates $\chi^{i}$:
\begin{eqnarray}
ds^{2} = d \chi^{i} \,  C_{ij} \,  d\chi^{j} \, \, .
\end{eqnarray}
In conclusion, it has been shown that the action (\ref{act}) describes a mixture of $D$ chiral scalars  $\chi_{+}^{\mu}$  and $D$ antichiral scalars $\chi_{-}^{\mu}$ ($\mu = 1, \dots, D$), which can be regarded as the components of the $2D$-dimensional vector $\chi^{i} \equiv (\chi_{-}^{\mu},  \chi_{+}^{\mu})$,  with $i=1, \dots, 2D$.

In  the action (\ref{actchir})  the ``non-chiral" basis of fields ${\cal X}^{i}\equiv(X^{\mu}, \tilde{X}_{\mu}$) can be introduced, with
\begin{eqnarray}
X^{\mu}\equiv \frac{1}{\sqrt{2}} ( \chi_{+}^{\mu}+\chi_{-}^{\mu})~~;~~\tilde{X}_{\mu}\equiv \frac{1}{\sqrt{2}} \delta_{\mu \nu} ( \chi_{+}^{\nu}-\chi_{-}^{\nu}),
\end{eqnarray}
in which the matrix $C$ becomes off-diagonal:
\begin{eqnarray}
C_{ij}=-\Omega_{ij}~~;~~\Omega_{ij}=\left(\begin{array}{cc}
           0_{\mu \nu} & \mathbb{I}^{ \,\, \nu}_{\mu} \\
           \mathbb{I}^{\mu}_{\,\, \nu}&0^{\mu \nu}\end{array}\right)\label{34} \, ,
\end{eqnarray}
with $(\Omega)_{ij}=(\Omega^{-1})^{ij}$.  The condition (\ref{2}) becomes the constraint $M^{-1}~=\Omega^{-1} M\Omega^{-1}$ on the  symmetric matrix $M$ that has $D^{2}=D(D+1)/2+D(D-1)/2 $  independent elements and, thus, it can be  parametrized by a symmetric matrix $G$ and  an antisymmetric one $B$.  The  expression for $M$, defined up to a sign, being the above constraint quadratic in it, is:
\begin{eqnarray} \label{M_ij}
 M_{ij}= \pm\left( \begin{array}{cc}
                                       (G-B\,G^{-1}B)_{\mu \nu} & (B\, G^{-1})_{\mu}^{\,\, \nu}\\
                                       (-G^{-1}\, B)^{\mu}_{\,\, \nu}  & (G^{-1})^{\mu \nu} \end{array}\right)  \,\, . \label{35}
\end{eqnarray}
The matrix $M$ is the so-called {\em generalized metric} \cite{TPL, TNP, 0605149, 1006.4823}. At the end of this section, it will be observed that only the positive sign of $M$  determines a positive definite Hamiltonian. Hence, $M$   is considered positive in eq. (\ref{35}).

In the non-chiral basis the action (\ref{act}) can be expressed as:
\begin{eqnarray}
S=\frac{1}{2} \int d^{2} \xi \,\, e \left[ \Omega_{ij} \nabla_{0} {\cal \chi}^{i} \nabla_{1} {\cal \chi}^{j} - M_{ij}  \nabla_{1} {\cal \chi}^{i} \nabla_{1} {\cal \chi}^{j} \right]. \label{actics}
\end{eqnarray}
It is invariant under the $O(D,D)$ transformations:
\begin{eqnarray}
{\cal \chi}'={\cal R} {\cal \chi}~~;~~M'={\cal R}^{-t}M {\cal R}^{-1}~~;~~{\cal R}^{t}\Omega {\cal R} =\Omega~~;~~{\cal R} \in O(D,D) \,\, \label{40}
\end{eqnarray}
showing that the background itself suitably transforms. One can immediately see that the matrix $\Omega$ belongs to  $O(D,D)$ and, in particular, when ${\cal R}^i_{~j}=\Omega_{ij}$, the action (\ref{actics}), expressed in terms of $X^{\mu}$ and $\tilde{X}_{\mu}$
\begin{eqnarray}
S= \frac{1}{2} \int d^{2} \xi e \left[ \nabla_{0} X^{\mu} \nabla_{1} \tilde{X}_{\mu} + \nabla_{0} \tilde{X}_{\mu} \nabla_{1} X^{\mu} - (G-B\,G^{-1}B)_{\mu \nu} \nabla_{1} X^{\mu} \nabla_{1} X^{\nu} \right. \nonumber \\
 \,\,\,\,\,\,\,\,\, \left.  - \,\, (B\, G^{-1})_{\mu}^{\,\, \nu} \nabla_{1} X^{\mu} \nabla_{1} \tilde{X}_{\nu} +   (G^{-1}\, B)^{\mu}_{\,\, \nu} \nabla_{1} \tilde{X}_{\mu} \nabla_{1} X^{\nu} - (G^{-1})^{\mu \nu} \nabla_{1} \tilde{X}_{\mu} \nabla_{1} \tilde{X}_{\nu} \right]
\end{eqnarray}
exhibits what in string theory will become the more familiar T-duality invariance under $X \leftrightarrow \tilde{X}$ with a consequent transformation of the generalized metric given by $M'=M^{-1}$.

Hence, once can claim that the sigma-model action (\ref{act}), even if non-covariant,  is the candidate to describe a bosonic string in the background constituted by $G$ and $B$ compactified on a torus $T^{D}$. It exhibits a manifest T-duality invariance $O(D,D)$. So one can introduce the string tension $T$ that makes $S$ dimensionless (in natural units) with the fields $\chi^{i}$ interpreted as the string coordinates on the double torus $T^{2D}$:
\begin{eqnarray}
S= - \frac{T}{2} \int d^{2} \xi \, e \left[ C_{ij} \nabla_{0} \chi^{i} \, \nabla_{1} \chi^{j}  + M_{i j} \nabla_{1} \chi^{i} \nabla_{1} \chi^{j} \right].  \label{actstr}
\end{eqnarray}
The string tension  $T$ can be, as usual, expressed in terms of $l$, the fundamental length of the theory, through the relation $T=1 / (2 \pi l^2)$. It is to be observed here that eqs. (\ref{fi0}) and (\ref{2}) can be recast in the following covariant form:
\begin{eqnarray}
-\epsilon_{ab}C_{ij} \partial^b \chi^{j} + M_{ij} \partial_a \chi^{j}=0 \, . \label{constr}
\end{eqnarray}
It will be shown in the following that the two equations in (\ref{constr}) coincide with the constraints imposed in the covariant formulation of the manifestly T-dual invariant bosonic string theory. In this case, their role is to keep only the physical degrees of freedom.

The double torus $T^{2D}$ that is going to be considered now is defined by the identification ${\cal X} \equiv {\cal X}+ 2 \pi l {\cal L}$, being ${\cal L} = (w, lp)$ a vector spanning a Lorentzian lattice $\Lambda^{D,D}$. In components, the identification becomes:
\begin{eqnarray}
X^{\mu}(\tau,\sigma + \pi) = X^{\mu}(\tau, \sigma) +2\,\pi\, l\,  w^{\mu}~~;~~\tilde{X}_{\mu}(\tau, \sigma + \pi) = \tilde{X}_{\mu} (\tau, \sigma) +2\pi\, l^{2} \,  p_{\mu} .\label{t33}
\end{eqnarray}
On the torus the previous symmetry $O(D,D; {\mathbb R})$ is broken to its discrete subgroup $O(D,D; {\mathbb Z})$.

In order to reconduce the action (\ref{act}) to a sum of Floreanini-Jackiw  Lagrangians, it is necessary to put the matrices $C$ and $M$ simultaneously in a block-diagonal form. This is performed by the matrix
\begin{eqnarray}
({\cal T}^{-1})^{ij} = \frac{1}{\sqrt 2} \left(\begin{array}{cc}
                                               (G^{-1})^{\mu \nu} & (G^{-1})^{\mu \nu} \\
                                               (-E^{t} \,G^{-1})_{\mu}^{\,\, \nu} & (E\, G^{-1})_{\mu}^{\,\, \nu} \end{array}\right) \, , \label{calT}
\end{eqnarray}
where $E \equiv G + B$. In fact, the matrix ${\cal T}^{-1}$ transforms $C$ and $M$ respectively into
\begin{eqnarray}
{\cal T}^{-t} C {\cal T}^{-1}=  \left(\begin{array}{cc}
                       G^{-1}&0\\
                       0&-G^{-1}\end{array}\right)\equiv  {\cal C}^{-1} ~~; ~~{\cal T}^{-t}M {\cal T}^{-1}=  \left(\begin{array}{cc}
                       G^{-1}&0\\
                       0&G^{-1}\end{array}\right)\equiv {\cal G}^{-1}\label{trgo}
\end{eqnarray}
and introduces new coordinates $\Phi_{i} = {\cal T}_{ij} {\cal X}^{j} \equiv (X_{R\,\mu} , X_{L\, \mu})$,  in terms of which the $R$ and $L$ sectors  are completely decoupled also in the presence of the $B$-field. The matrix ${\cal G}^{-1} $ is the generalized metric in the chiral coordinates system.

The matrix ${\cal T}$ is not an element of  the group $O(D,D)$ because it changes the metric $C$ in ${\cal C}^{-1}$. It has to be seen as leading to a field redefinition that makes the explicit dependence on the $B$-field disappear in the action. An $O(D,D)$ transformation leaves invariant the metric ${\cal C}$ but, in general, transforms ${\cal G}^{-1}$ in a non-diagonal matrix, as shown in Appendix \ref{B}. Hence, such matrix, after the action of the non-compact group, will exhibit all the dependence on the fields $G$ and $B$ as any general symmetric $O(D,D)$ matrix.
The transformations which leave invariant the two metrics  ${\cal G}$ and ${\cal C}$, and hence the action,  belong to the subgroup $O(D)\times O(D)$ of the original orthogonal group $O(D,D)$.

In the flat gauge, previously introduced, the action becomes:
\begin{eqnarray}
S \equiv \int d^{2} \xi [ {\cal L}_{R} + {\cal L}_{L} ] \, , \label{actrl}
\end{eqnarray}
with
\begin{eqnarray}
\frac{1}{T}\, {\cal L}_{L; R} \equiv \pm \frac{1}{2}  \partial_{0} X_{L; R}^{t} G^{-1} \partial_{1} X_{L; R}   - \frac{1}{2} \partial_{1} X_{L; R}^{t} G^{-1} \partial_{1} X_{L; R} \label{FJrl}
\end{eqnarray}
which is just the realization in the double string theory of the Floreanini-Jackiw Lagrangians (\ref{FJL}) with a non-vanishing Kalb-Ramond field as background. Eq. (\ref{constr}) can be rewritten in a more compact  form in terms of the Hodge duals of $dX_{R}$ and $dX_{L}$\footnote{The conventions used here for $p$-forms in a $D$-dimensional space-time with metric $G$ having signature $(-,+^{(D-1)})$ are the following: $w_{(n)}=\frac{1}{n} w_{\mu_{1} \dots \mu_{n}} dx^{\mu_{1}} \wedge \dots \wedge dx^{\mu_{n}}$ and $*w_{(n)}=\frac{\sqrt{-\mbox{det}G}}{n! (D-n)!} \epsilon_{\nu_{1} \dots \nu_{D-n} \mu_{1} \dots \mu_{n}} w^{\mu_{1} \dots \mu_{n}} dx^{\nu_{1}} \wedge \dots \wedge dx^{\nu_{n}} $ with $\epsilon^{0 1 \dots (D-1)}=1$.} as:
\begin{eqnarray}
*dX_R=dX_R ~~;~~*dX_L=-dX_L \,\, . \label{cfet}
\end{eqnarray}
The next aim is to solve the self- and anti-self-dual conditions (\ref{cfet}) with the boundary conditions already given but rewritten in the new chiral basis. It is worth to observe here that this corresponds to solve both the equations of motion for the string coordinates and the constraint $\epsilon^{ab}t_{ab}=0$, necessary to recover the local Lorentz invariance. Hence, along the solution, only two conditions derive from the original constraints $t_{ab}=0$.

The solution of the duality equations (\ref{cfet}), with identifications on the torus now rewritten as:
\begin{eqnarray}
X_{R \, \mu}[\tau - (\sigma +\pi)] = X_{R \, \mu}(\tau - \sigma) -2\pi\, l^{2}\,  p_{R \, \mu} \label{cr} \\
X_{L \, \mu}[\tau + (\sigma +\pi)]= X_{L \, \mu}(\tau + \sigma) +2\pi\, l^{2}\,  p_{L \, \mu} \label{cl}
\end{eqnarray}
with
\begin{eqnarray}
\left(\begin{array}{c} -l p_R\\ l p_L\end{array}\right)={\cal T} \left(\begin{array}{c}w\\ lp\end{array}\right)\,\,,\label{nl}
\end{eqnarray}
 is given by:
\begin{eqnarray}
X_R (\tau - \sigma)&=& x_R+2\,l^2\,p_R(\tau -\sigma)+i l\sum_{n\neq 0}\frac{{\alpha}_n}{n}e^{-2i n(\tau- \sigma)} \label{xr_exp} \\
 X_L (\tau + \sigma) &=& x_L+2\,l^2\,p_L(\tau +\sigma)+i l\sum_{n\neq 0}\frac{\tilde{\alpha}_n}{n}e^{-2i n(\tau+\sigma) } \label{xl_exp}
\end{eqnarray}
formally identical to the usual expansion of the right and left bosonic string coordinates.

The relation between $(X_{R}, X_{L})$ and $(X, \tilde{X})$ implies:
\begin{eqnarray}
X (\tau, \sigma) & = & x + 2l^2\,G^{-1} \left[ p -  B \frac{w}{l} \right] \tau + 2 l w\sigma \label{seqx} \\
& & +  \frac{i l}{\sqrt 2} \,G^{-1}\sum_{n\neq 0} \frac{e^{-2in\tau}}{n} \left[ {\alpha}_n e^{+2in\sigma}+\tilde{\alpha}_n e^{-2 i n\sigma}\right] \nonumber
\end{eqnarray}
and
\begin{eqnarray}
\tilde{X}(\tau, \sigma) &=& \tilde{x}  + 2l^2 \left[ BG^{-1} p + (G - B G^{-1} B) \frac{w}{l} \right] \tau
+ 2 l^2 p\sigma \label{seqx1} \\
&& + \frac{i l}{\sqrt 2} \sum_{n\neq0} \frac{e^{-2in\tau}}{n} \left[ - E^t G^{-1}{\alpha}_n e^{+2in\sigma} + E G^{-1} \tilde{\alpha}_n e^{-2in\sigma}\right] \nonumber
\end{eqnarray}
where $x$ and $\tilde{x}$ are defined by:
\begin{eqnarray}
x = \frac{1}{\sqrt{2}} G^{-1} ( x_R + x_L ) ~~;~~ \tilde{x} = \frac{1}{\sqrt{2}} ( - E^t G^{-1} x_R + E G^{-1} x_L )
\end{eqnarray}
and from eq. (\ref{nl}):
\begin{eqnarray}
p_R= \frac{1}{\sqrt{2}} \left[p - E \frac{w}{l} \right] ~~;~~ p_L=
 \frac{1}{\sqrt 2}\left[p+ E^t\frac{w}{l}\right]  \,\, .
\end{eqnarray}
Reading $p$ and $w$ respectively as a momentum and a winding number, one can see that these expressions are the same as the ones holding in the usual closed string compactified on a torus.

The Hamiltonian of this system turns out to be:
\begin{eqnarray}
H= \frac{T}{2}\int_0^\pi d \sigma ~ \partial_{1}  \Phi^{t} \, {\cal G}^{-1} \, \partial_{1}  \Phi \,\, .
\end{eqnarray}
Having chosen for $M$ the positive sign, $H$ is positive definite.

It is convenient to introduce the world-sheet light-cone coordinates $\sigma^+=\tau+\sigma$ and $\sigma^-=\tau-\sigma$. In terms of these ones,  the components of the $t$-tensor turn out to be:
\begin{eqnarray}
t_{++} &=&  \partial_+X_R^tG^{-1} \partial_+ X_R + \partial_+X_L^t G^{-1} \partial_+X_L
- 2 \partial_+X_L^tG^{-1} \partial_-X_L \nonumber \\
&&   \label{12cos} \\
t_{--} &=&  \partial_-X_R^tG^{-1} \partial_- X_R + \partial_- X_L^t G^{-1} \partial_- X_L
- 2 \partial_+X_R^tG^{-1} \partial_-X_R \nonumber
\end{eqnarray}
while the Weyl invariance imposes $t_{+-} = -t_{-+}$, with
\begin{eqnarray}
t_{+-} = - \frac{1}{4} \epsilon^{ab} t_{ab} =\partial_- X_L^t G^{-1} \partial_- X_L - \partial_+ X_R^t G^{-1} \partial_+X_R   \label{3cos}
\end{eqnarray}
and $\partial_{\pm} = \frac{1}{2} (\partial_{0} \pm \partial_{1})$. The quantity defined in (\ref{3cos}) is of course vanishing on-shell, while the other two quantites in (\ref{12cos}) have to be seen as contraints to be imposed at the classical and quantum level. On-shell they look like the contraints on $T_{++}$ and $T_{--}$ for the energy-momentum tensor in the usual bosonic string theory leading to the Virasoro algebra.

\subsection{Analysis of the constraints and quantization}

The quantization of two-dimensional self- and anti-self-dual fields has  been extensively investigated in the literature \cite{TPL, TNP, FJ, 0910.0431}. It is already known, for example,  that these systems are characterized by  primary second class constraints  which require the introduction of Dirac brackets.  The action in exam is the one in eq. (\ref{actrl}). It describes the dynamics of $D$ chiral and $D$ antichiral scalar fields.

Since the Lagrangians are linear in the time derivative of the fields, the conjugate momenta
\begin{eqnarray}
P_R \equiv \frac{\partial {\cal L}_R}{\partial (\partial_0 X_R^t)} = - \frac{T}{2} G^{-1} \partial_1 X_R ~~;~~ P_L \equiv \frac{\partial {\cal L}_L}{\partial (\partial_0 X_L^t)} = \frac{T}{2} G^{-1} \partial_1 X_L \label{defm}
\end{eqnarray}
define the primary constraints of the theory:
\begin{eqnarray}
\Psi_R(P_R,\, X_R)=P_R + \frac{T}{2} G^{-1} \partial_1X_R\approx 0 ~~;~~ \Psi_L(P_L,\,X_L)=P_L- \frac{T}{2} G^{-1} \partial_1X_L\approx 0 \,\, . \label{clcos}
\end{eqnarray}
The classical dynamics of the system is studied by defining the Poisson brackets
\begin{eqnarray}
\left\{P_{R; L}(\tau,\,\sigma),\,X_{R; L}^{t}(\tau,\,\sigma')\right\}_{PB}= \mathbb{I} \, \delta(\sigma-\sigma') \label{pb} \, .
\end{eqnarray}
According to the previous definition, the primary constraints satisfy the following equal `time' algebra
\begin{eqnarray}
\left\{ \Psi_{R; L}(\tau,\sigma),\,\Psi_{R; L}^{t}(\tau, \sigma')\right\}_{PB} = \mp TG^{-1}\delta'(\sigma-\sigma') \, , \label{eq51}
\end{eqnarray}
with $\delta'(x)=\partial_x\delta(x)$ and the upper [lower] sign on the right hand side of the previous identity refers to the label $R$ [$L$] on the left  of the same  equation. The algebra in eq. (\ref{eq51}) implies that these primary constraints are second class.

As it has been shown, further constraints hold in the theory, i.e. $t_{ab} = 0$. A rigorous analysis of all the constraints requires the study of the complete algebra generated by all of them.

By analogy with the standard procedure followed in string theory, the constraints are evaluated here on the solution of the equation of motion for the fields $X_{R;\, L}$. One of the constraints, $t_{+-} \approx 0$, is already satisfied on it. The other constraints become:
\begin{eqnarray}
\Psi_R=P_R - \frac{T}{2} G^{-1} \partial_{-} X_R\approx 0 ~~;~~ \Psi_L=P_L- \frac{T}{2} G^{-1} \partial_{+}X_L\approx 0 \,\,
\end{eqnarray}
and
\begin{eqnarray}
 t_{++} &=& \, \partial_+X_L^t G^{-1}\partial_+X_L\approx 0 \nonumber\\
\\
t_{--} &=&  \partial_-X_R^t\,G^{-1}\partial_-X_R \approx 0 . \nonumber
\end{eqnarray}
On the equations of motion, the algebra of the constraints reads:
\begin{eqnarray}
\left\{ \Psi_{R} (\tau, \sigma), t_{--}(\tau, \sigma') \right\}_{PB}= \delta'(\sigma - \sigma') G^{-1} \partial_{-} X_{R}(\tau -\sigma) \approx 0
\end{eqnarray}
(with a similar expression for $\Psi_{L}$ and $t_{++}$).  Here the last relation  comes from the constraint $t_{--} \approx 0$.

As already stressed, according to the Dirac analysis, the presence of second class constraints
leads to the introduction of the Dirac brackets. In Appendix A their definition is explicitly given. A straightforward computation leads to:
\begin{eqnarray}
\left\{X_{R; L}(\tau,\,\sigma),\,X^{t}_{R; L}(\tau,\,\sigma')\right\}_{DB}&=&\mp \frac{G}{T}\epsilon(\sigma-\sigma')   \nonumber  \\
\left\{P_{R; L}(\tau,\,\sigma),\,X^{t}_{R; L}(\tau,\,\sigma')\right\}_{DB}&=&\,\,\,\,\,\frac{1}{2} {\mathbb I} \, \delta(\sigma-\sigma') \label{RLDB}\\
\left\{P_{R; L}(\tau,\,\sigma),\,P^{t}_{R; L}(\tau,\,\sigma')\right\}_{DB}&=&\pm\frac{T}{4}G^{-1}\delta'(\sigma-\sigma') \nonumber
\end{eqnarray}
where $\epsilon(\sigma-\sigma')$ is the step function defined in Appendix A.

It is also useful to give the equal time Dirac brackets of the original variables $X$ and $\tilde{X}$:
\begin{eqnarray}
&&\left\{X(\tau,\,\sigma),\,\tilde{X}^{t}(\tau,\,\sigma')\right\}_{DB}=\frac{1}{T}{\mathbb I}\, \epsilon(\sigma-\sigma') \nonumber\\
&&\left\{{P}(\tau,\,\sigma),\,{X}^{t}(\tau,\,\sigma')\right\}_{DB}=\left\{\tilde{P}(\tau,\,\sigma),\,
\tilde{X}^{t}(\tau,\,\sigma')\right\}_{DB}=
\frac{1}{2}{\mathbb I} \, \delta(\sigma-\sigma') \label{RLDB1}
\\
&&\left\{{P}(\tau,\,\sigma),\,{\tilde P}^{t}(\tau,\,\sigma')\right\}_{DB}= -\frac{T}{4}{\mathbb I} \, \delta'(\sigma-\sigma') \nonumber
\end{eqnarray}
being $ P$ and $\tilde{P}$ the conjugate momenta with respect to $X$ and $\tilde{X}$.

The double world-sheet sigma-model is now quantized by replacing the Dirac brackets with the corresponding commutator according to the well-known substitution:
\begin{eqnarray}
\left\{ \cdot \, , \, \cdot \right\}_{DB} \rightarrow -i[ \cdot \, , \, \cdot] \label{sub} \,\, .
\end{eqnarray}

The Dirac brackets of second class constraints with themselves and with  any function defined on the phase space are vanishing. At the quantum level, this means that they commute with any operator and therefore they can be considered as $c$-numbers \cite{Dirac} having to be zero.  Hence, at the quantum level, eqs. (\ref{clcos}) are operator identities that can be ``strongly" put to zero. One can then write on-shell:
\begin{eqnarray}
P_{R}=TG^{-1}\left[  l^2 p_R + l \sum_{n\neq0} e^{-2in(\tau-\sigma)} \alpha_n\right]   ~;~
P_{L}=TG^{-1}\left[  l^2 p_L + l \sum_{n\neq0} e^{-2in(\tau+\sigma)} \tilde{\alpha}_n\right]  \, .\nonumber
\end{eqnarray}
 The Dirac brackets given in eqs. (\ref{RLDB}), via the usual substitution in eq. (\ref{sub}),
determine the following commutators for the Fourier modes:
\begin{eqnarray}
[p_{R; L},\,x_{R; L}^t]= i G ~~;~~ [\alpha_m,\, \alpha_n^t]= m G\delta_{m+n}~~;~~[\tilde{\alpha}_m,\,\tilde{\alpha}_n^t]= m G\delta_{m+n}  \, . \label{fmcr}
\end{eqnarray}
Details about the previous identities are given in Appendix \ref{B}.

The constraints involving the Laurent expansions of the components $t_{++}$ and $t_{--}$ are:
\begin{eqnarray}
 t_{++} &=&  \partial_+X_L^t G^{-1}\partial_+X_L\, \equiv \, \frac{4}{\pi T}\sum_{n\in \mathbb{Z}} \tilde{L}_n\,e^{-2in(\tau+\sigma)}=0 \\
 t_{--} &=&  \partial_-X_R^t\,G^{-1}\partial_-X_R\equiv \frac{4}{\pi T}\sum_{n\in \mathbb{Z}} {L}_n\,e^{-2in(\tau-\sigma)}=0 \, ,
\end{eqnarray}
where
\begin{eqnarray}
\tilde{L}_n = \, \frac{T}{4} \int_0^\pi d\sigma \,\, e^{2in \sigma} \partial_+X_L^t\,G^{-1}\partial_+X_L \, = \frac{1}{2}\sum_{m\in\mathbb{Z}} \tilde{\alpha}_m^t\,G^{-1}\, \tilde{\alpha}_{n-m}-a\delta_{n,0}  \,\, \\
{L}_n = \frac{T}{4} \int_0^\pi d\sigma e^{-2in\sigma} \partial_-X_R^t \,G^{-1}\partial_-X_R = \frac{1}{2}\sum_{m\in\mathbb{Z}}{\alpha}_m^t \,G^{-1}\, {\alpha}_{n-m}-a\delta_{n,0} \, .
\end{eqnarray}
Here,  $\tilde{\alpha}_0 \equiv lp_L$ and $\alpha_0 \equiv lp_R$  have been defined and, by analogy with the usual Virasoro generators,  a constant $a$ has been added in the zero components of the {\em Virasoro-like} generators in order to take into account the normal ordering ambiguity.

Finally, one observes that the following relation between the Hamiltonian and the components of the $t$-tensor holds on-shell:
\begin{eqnarray}
\frac{H}{2} = \frac{T}{4}\int_0^\pi d\sigma \left[t_{++}+t_{--}\right]=  \tilde{L}_0+L_0 \,\, .
\end{eqnarray}
Again, it is the generalization, in this context, of the usual relation between the Hamiltonian and the Virasoro generators.

In this section, similarities and differences between the ordinary bosonic string and the double string theory have emerged out. Among the former, the most relevant are given by the coincidence of  eqs. (\ref{fmcr})  with the ones usually satisfied by the Fourier modes of the string coordinates in bosonic string theory and by the fact that the Virasoro-like generators, once expressed in terms of their Fourier modes, are formally identical to the standard Virasoro generators. Hence, the quantum  anomaly both in the sectors $\alpha^{\mu}_{n}$ and $\tilde{\alpha}^{\mu}_{n}$ is vanishing with $\mu$ varying in  $26$ space-time dimensions. Of course, this critical dimension is now  equal to the sum of the number of the non-compact dimensions and of the $D$ compact dimensions of the torus $T^{D}$.

Furthermore, it is worth to observe here that  the free double string theory has to be considered as an extension of the usual bosonic string theory. Indeed, as already stressed in the original paper by Tseytlin\cite{TNP}, in the free double string theory it is always possible to integrate out the $\tilde{X}$ coordinate and, modulo boundary terms which have to be carefully treated, one can always recover the action of the usual string theory. However, the main difference between the two formulations, also in the free case, is the presence of the zero mode $\tilde{x}$ of the dual coordinate $\tilde{X}$ which turns out to be completely independent on the zero mode $x$  of the field $X$. This  feature  allows to introduce two completely independent  and decoupled $R$ and $L$ sectors,  when  the $B$-field is in the background.

\section{The covariant double string sigma-model}  \label{hull}

In this section, attention will be focused on the Lorentz and $O(D,D;\mathbb{Z})$ manifestly invariant  formulation of the double string theory by Hull \cite{0406102} and how it is related to the non-covariant action proposed by Tseytlin\cite{TPL,TNP}.

In the covariant approach,  the starting point is the sigma-model defined by the coordinates $(Y(\tau,\,\sigma),\,\cal{X}(\tau,\,\sigma))$  mapping the string world-sheet in the target space. Locally, the target space looks like $\mathbb{R}^{1,  d-1}\otimes T^{2D}$ where the coordinates $Y \equiv (Y^{I})$ , $I=0, \dots, d-1$ are associated with the non-compact space-time  while the coordinates ${\cal X}\equiv ({\cal X}^i)$, $i=1,\dots, 2D$, through the identification given in eq. (\ref{t33}), describe the double torus.
The world-sheet action proposed in Ref.\cite{0406102} is
\begin{eqnarray}
S=- \frac{T}{4} \int  d{\cal X}^i\,M_{ij}(Y)\wedge*d{\cal X}^j
\label{inva}
\end{eqnarray}
where $M$ 
is 
a generalized metric. 

The action, supplemented by the torus identifications given in eq. (\ref{t33}), is invariant under the   $GL(2D;\mathbb{Z})$ group which is the manifest symmetry of the theory\cite{0406102}. Since the number of the coordinates on the torus has been doubled, a self-duality constraint that could halve them has to be imposed:
\begin{eqnarray}
*M_{ij}\,d{\cal X}^j =  -\Omega_{ij}\, d{\cal X}^j\label{Hcos}\;\; . \label{gugo}
\end{eqnarray}
Here $\Omega$ is the $O(D,D)$ invariant metric defined in eq. (\ref{34}). With this choice, the invariance of the theory reduces to the one under  $O(D,D; \mathbb{Z})$.  Eq. (\ref{gugo}) is identical to the $\epsilon$-trace constraint of the Tseytlin action necessary for restoring, in that case, the Lorentz local invariance.

The energy-momentum tensor obtained from this action turns out to be:
\begin{eqnarray}
T_{\alpha\beta}=-\frac{4}{T } \frac{1}{\sqrt {- g}} \frac{\delta S}{\delta g^{\alpha\beta}}=
\partial_{\alpha} {\cal \chi}^t
\,M\,\partial_\beta {\cal \chi} -\frac{1}{2} g_{\alpha\beta}\partial_\gamma{\cal \chi}^t\,M\,
\partial^{\gamma}{\cal \chi}   .
\end{eqnarray}
It is traceless because  of the Weyl invariance.  The latter, together with the invariance under reparametrizations of the world-sheet, is used to gauge-fix  the two-dimensional metric so that $g_{\alpha \beta} =\eta_{\alpha\beta}$.

The equations of motion for ${\cal \chi}$, clearly satisfied on the constraint surface, are:
\begin{eqnarray}
d*(Md{\cal \chi})=0
\label{eqmot1}
\end{eqnarray}
with boundary conditions given by the surface integral:
\begin{eqnarray}
-\frac{T}{2} \left. \int d\tau \, \delta {\cal X}^{t} M \partial_{1} {\cal X} \right|^{\sigma = \pi}_{\sigma=0}
\end{eqnarray}
vanishing if periodicity conditions, peculiar of closed strings, are imposed.

By proceeding in analogy with the non-covariant formulation,
it is convenient to introduce the right and left coordinates $\Phi_i=(X_{R \, \mu},\,X_{L\, \mu})$:
\begin{eqnarray}
\Phi_i= {\cal T}_{ij}{\cal \chi}^j~~;~~
{\cal T} = \frac{1}{\sqrt 2} \left(\begin{array}{cc}
                                               E& \mathbb{-I}\\
                                               E^t&-\mathbb{+I}\end{array}\right) \,\, .
\end{eqnarray}
It has been already shown in eq. (\ref{trgo}) how the matrix ${\cal T}$ acts on  the generalized metric and on  the $O(D,D)$ invariant  one.
According to those transformations, in this new system of coordinates the matrix ${\cal C}^{-1}$
plays the role of the $O(D,D)$ invariant metric and ${\cal G}^{-1}$ the one of generalized metric.

The action (\ref{inva}), when rewritten in terms of these coordinates, becomes:
\begin{eqnarray}
S= -\frac{T}{4} \int  d\Phi ^t\,{\cal G}^{-1}\wedge*d\Phi \,\, .
\end{eqnarray}
It is worth to observe that in this frame any dependence on the Kalb-Ramond field disappears  making the quantization of the theory quite simple and  transparent.

The energy-momentum tensor can be equivalently written as:
\begin{eqnarray}
T_{\alpha\beta}= \partial_{\alpha}\Phi^t{\cal G}^{-1} \partial_{\beta}\Phi -\frac{1}{2} \eta_{\alpha\beta} \partial^\gamma \Phi^t{\cal G}^{-1} \partial_\gamma\Phi~
\end{eqnarray}
and the conjugate momentum is:
\begin{eqnarray}
{\cal P}\equiv\left( \begin{array}{c}
 P_R(\tau,\,\sigma)\\ P_L (\tau,\,\sigma)\end{array}\right)= \frac{T}{2}{\cal G}^{-1}\partial_0\Phi \,\, .
\end{eqnarray}
The Hamiltonian turns out to be:
\begin{eqnarray}
H=\frac{T}{4}\int^\pi_0d\sigma\left[ \partial_0\Phi^t{\cal G}^{-1}\partial_0\Phi+\partial_1\Phi^t{\cal G}^{-1}\partial_1\Phi\right] \,\, .
\end{eqnarray}
In the new basis the constraints become the ``duality" conditions
\begin{eqnarray}
\frac{2}{T}\,{\Psi}_{R} \equiv dX_R- *dX_R=0~;~~ \frac{2}{T}{\Psi}_{L} \equiv dX_L+*dX_L=0  \label{Hncos}
\end{eqnarray}
that generalize to this case the self-dual and anti-self dual constraints satisfied by the usual string coordinates compactified on a torus. Eqs. (\ref{Hncos}) formally determine four conditions for the $X_{R;L}$ coordinates. However, only two of them are independent and they  can be written in the following form:
\begin{eqnarray}
\left( {\Psi}_{R;L} \right)_{0} = \pm\left( {\Psi}_{R;L} \right)_{1} \equiv T\partial_\pm X_{R;L}= G P_{R;L}\pm\frac{T}{2} \partial_{1} X_{R;L} =0
\end{eqnarray}
where the  definition of the conjugate momentum has been used. These constraints coincide with the second-class ones in eq.  (\ref{clcos}) and so satisfy the algebra given in eq. (\ref{eq51}), behaving like second-class constraints.

The identities given in eq. (\ref{Hncos})  can be incorporated in the action\footnote{One of the authors, F. P., is deeply grateful to Dmitri Sorokin for a very helpful discussion on this topic.}, according to the procedure defined in Refs. \cite{PST1,PST2} (see also Refs. \cite{1207.6110, 1308.4418}). Following this procedure,  the self- and anti-self-dual conditions can be taken into account by introducing an auxiliary one-form $u$ and by writing
\begin{eqnarray}
S &=& -\frac{T}{4} \int d\Phi ^t\,{\cal G}^{-1}\wedge*d\Phi \,+\, \frac{1}{T}\int d^2\sigma \frac{1}{u^2}\, u^\alpha\, \Psi_\alpha^t\,{\cal G}^{-1}\,\Psi_\beta \,u^\beta \, , \label{85}
\end{eqnarray}
being $\Psi\equiv(\Psi_R \, , \,\Psi_L)$ and $u_\alpha=\partial_\alpha \,a$ with $a$ an auxiliary scalar field. The action (\ref{85}) is invariant under the following local transformations:
\begin{eqnarray}
\delta a= \varphi ~~;~~\delta\Phi= \frac{2}{T} \,\frac{\varphi\,   u_{\alpha}\Psi^\alpha}{u^2}  \,\, .
\end{eqnarray}

The symmetries of this action allow to choose the gauge $u_\alpha=\delta_{\alpha}^0$\cite{PST2} with $u^2 = u_{\alpha}u^{\alpha}=-1$ and, in this gauge, the previous action coincides with the one  written in eq. (\ref{actrl}) showing the equivalence between  the  constrained theory  by Hull and the one by Tseytlin. The proof of the equivalence completely fixes the relative  overall coefficients of the two actions.

The  chosen gauge breaks the Lorentz invariance of the original action. However, there exists a linear combination of Lorentz and gauge transformations, which preserves  the choice $u_\alpha=\delta_\alpha^0$. This transformation is fixed by requiring  $ \delta u^\alpha=
v\epsilon^{\alpha\beta}\delta_\beta^0+\partial^\alpha\varphi=0$, being the first term an infinitesimal Lorentz transformation with constant parameter $v$ and the second one an infinitesimal gauge rotation. This equation implies $\varphi =v\sigma$. The Lorentz transformations of the field $\Phi$ are now replaced by \cite{PST2}:
\begin{eqnarray}
\delta \Phi =v\xi_\alpha \epsilon^{\alpha\beta}\partial_\beta\Phi +\frac{2v\sigma}{T}\frac{u_{\alpha} \,\Psi^\alpha}{u^2} \,\, .
\end{eqnarray}

In the following discussion, instead of implementing constraints in the action, it will be preferred to perform the Dirac analysis of  the constrained systems.

It is convenient, in analogy with string theory, to introduce the world-sheet light-cone coordinates $\sigma^{\pm}=\tau\pm \sigma$. According to the standard rules of the tensor analysis, the non-vanishing components of the energy-momentum tensor in these coordinates are:
\begin{eqnarray}
T_{++}=\frac{1}{2}\left( T_{00}+T_{01}\right)= \partial_+\Phi^t{\cal G}^{-1} \partial_+\Phi~~;~~
T_{--}=\frac{1}{2}\left( T_{00}-T_{01}\right)= \partial_-\Phi^t{\cal G}^{-1}\partial_-\Phi \,\, ,
\end{eqnarray}
being, as usual, $\partial_\pm=\frac{1}{2}(\partial_0\pm\partial_1)$. It is also useful to express  the
components of the energy-momentum tensor in terms of the ``second class'' constraints:
\begin{eqnarray}
T_{++} & = & \frac{1}{T^2}\Psi^{t}_RG^{-1}\Psi_R+\partial_+X^{t}_LG^{-1}\partial_+X_L \nonumber \\
\\
T_{--} &= & \frac{1}{T^2} \Psi^{t}_LG^{-1}\Psi_L+\partial_-X^{t}_RG^{-1}\partial_-X_R . \nonumber
\end{eqnarray}
It is easy to check that the left and right sectors commute by definition, while
\begin{eqnarray}
\left\{T_{\pm \pm},\, \Psi_{R,L}\right\}_{PB} = \mp \frac{2}{T} \delta' (\sigma - \sigma') \Psi_{R,L} \approx 0
\end{eqnarray}
where the ``weak"  identity to zero is meant on the surface of the constraints. Furthermore, the following identity holds:
\begin{eqnarray}
 \left\{
\partial_{\mp}X_{R;L}(\tau,\,\sigma),\,\Psi_{R;L}(\tau,\,\sigma')\right\}=0 \,  .
\end{eqnarray}

The Hamiltonian in these coordinates becomes
\begin{eqnarray}
H & = & \frac{T}{2} \int_0^\pi d\sigma\left[ \frac{1}{T^2}\Psi^{t}_RG^{-1}\Psi_R +\partial_-X^{t}_RG^{-1}\partial_-X_R \right. \nonumber \\
 & & \qquad \qquad \left. + \frac{1}{T^2}\Psi^{t}_L G^{-1}\Psi_L +\partial_+X^{t}_LG^{-1}\partial_+X_L\right]
\end{eqnarray}
which has weakly vanishing Poisson brackets with the second class constraints.

Second class constraints are treated by the Dirac method of quantization\cite{Dirac}. This is also been done in the approach followed in Ref.\cite{0605114} (see also\cite{0605149}). Here,  the analysis is going to be extended  to the general torus $T^{D,D}$  also with a $B$-field background.
The Dirac brackets between the canonical coordinates are:
\begin{eqnarray}
\left\{ P_{R;L}(\tau,\,\sigma),\,X^{t}_{R;L}(\tau,\,\sigma')\right\}_{DB}&=&\,\,\,\,\, \frac{1}{2}\, {\mathbb I}\,  \delta(\sigma-\sigma')\nonumber\\
\left\{ X_{R;L}(\tau,\,\sigma),\,X^{t}_{R;L}(\tau,\,\sigma')\right\}_{DB}&=&\mp\frac{G}{T}  \epsilon(\sigma-\sigma') \label{DCB} \\
\left\{ P_{R;L}(\tau,\,\sigma),\,P^{t}_{R;L}(\tau,\,\sigma')\right\}_{DB}&=&\pm\frac{T}{4}G^{-1} \delta'(\sigma-\sigma') \nonumber \,\, .
\end{eqnarray}
The second class constraints can be now strongly imposed, yielding
$X_R\equiv X_R(\sigma^-)$ and $X_L\equiv X_L(\sigma^+)$. These identities, once solved with the closed string boundary conditions, lead to the the Fourier expansions  given in eqs. (\ref{xr_exp}, \ref{xl_exp}).

The expression of the energy-momentum tensor on the surface constraint simplifies becoming:
\begin{eqnarray}
T_{++}= \partial_+ X_L^t{ G}^{-1} \partial_+X_L~~;~~
T_{--}= \partial_-X^t_R{ G}^{-1} \partial_-X_R
\end{eqnarray}
while the Hamiltonian reduces to
\begin{eqnarray}
H=\frac{T}{2}\int_0^\pi d\sigma\left[\partial_-X^{t}_RG^{-1}\partial_-X_R+\partial_+X^{t}_LG^{-1}\partial_+X_L\right] \,\, .
\end{eqnarray}

Eq. (\ref{DCB}) determines the following Dirac brackets for  the coordinates Fourier modes:
\begin{eqnarray}
\{p_{R;L},\,x^{t}_{R;L}\}_{DB}= G ~;~\{ \alpha_m,\,\alpha^t_n \}_{DB}= i m \,G\,\delta_{m+n}~;~\{ \tilde{\alpha}_m,\,\tilde{\alpha}^t_n \}_{DB}=i m \,G\,\delta_{m+n} \, , \label{dirbra}
\end{eqnarray}
which again coincide with the Poisson brackets of the string modes in the bosonic string theory.

For completeness, it is interesting to give also the Dirac brackets among  the original coordinates $\chi$ and their momenta. In this frame the conjugate momentum is given by $P= {\cal T}^t {\cal P}$ and one has:
\begin{eqnarray}
\left\{{\cal P}(\tau,\,\sigma),\, \chi^t(\tau,\,\sigma')\right\}_{DB}&=& \,\,\,\,\, \frac{1}{2}\,\mathbb{I}\,\delta(\sigma-\sigma') \,\, \nonumber\\
\left\{{\chi}(\tau,\,\sigma),\, \chi^t(\tau,\,\sigma')\right\}_{DB}&=& \,\,\frac{\Omega^{-1}}{T} \, \epsilon(\sigma-\sigma') \,\, \\
\left\{{\cal P}(\tau,\,\sigma),\, {\cal P}^t(\tau,\,\sigma')\right\}_{DB}&=& - \frac{T}{4}\Omega \, \delta'(\sigma-\sigma') \nonumber  \,\, .
\end{eqnarray}
The previous Dirac brackets are invariant under $O(D,D; \mathbb{Z})$ transformations. This can be easily seen by observing that ${\cal P}'={\cal R}^{-t}{\cal P}$ and reminding that $\chi'={\cal R} \chi$.

The quantization of this theory is exactly the same as the Tseytlin one. It is trivially obtained by applying on eq. (\ref{dirbra}) the standard substitution given in eq. (\ref{sub}) which leads again to the eq. (\ref{fmcr}).

\section{Open string solutions} \label{openstrings}

The analysis performed so far is based on the mode expansion given in eqs. (\ref{seqx}) (\ref{seqx1}) which solve the duality constraints in (\ref{cfet}, \ref{Hncos})  with the boundary conditions in eq. (\ref{t33}). These are necessary to cancel out the surface integrals generated by the standard procedure used for the derivation of  equations of motion.

In order to explore the possibility to find open string like solutions of the duality equations,
it is useful to  write explicitly the boundary terms. In the Tseytlin and Hull models,  they are respectively equal to
\begin{equation}
\begin{array}{c}
\left. \left[ -\delta X_R^t G^{-1}\left( \partial_0+2\partial_1\right) X_R+\delta X_LG^{-1}\left( \partial_0 -2 \partial_1\right)X_L\right] \right|_{\sigma=0}^{\sigma=\pi}=0\\
\\
\left. \left[\delta X_R^tG^{-1}\partial_1X_R+\delta X_L^tG^{-1}\partial_1X_L\right] \right|_{\sigma=0}^{\sigma=\pi}=0 \, .
\end{array}
\end{equation}
By introducing the world-sheet light-cone coordinates $\sigma^{\pm}$ and  after some simple algebra, it is possible to write,  on-shell, both the boundary terms in the following form:
\begin{eqnarray}
\left.\left[ \delta X_R^t G^{-1}\partial_-X_R-\delta X_L^t G^{-1}\partial_+X_L\right] \right|_{\sigma=0}^{\sigma=\pi}=0 \,\, . \label{nbc0}
\end{eqnarray}
In the spirit of finding open string like solutions,   boundary conditions relating the $R$ and $L$ sectors  have to be imposed. Indeed the following identification \begin{eqnarray}
\left.\partial_- {X}_R (\tau-\sigma)\right|_{\sigma=0}^{\sigma=\pi}=\pm\left.\partial_+
{X}_L(\tau+\sigma)\right|_{\sigma=0}^{\sigma=\pi}\label{nbc1}~
\end{eqnarray} fulfills   eq. (\ref{nbc0})
since the expansions
\begin{eqnarray}
\left. \delta X_{R;L}(\tau\mp\sigma)\right|_{\sigma=0,\pi}=\left.\partial_{\mp}X_{R;L}\right|_{\sigma=0,\pi} \delta\tau
\end{eqnarray}
also determine $\delta X_R(\tau-\sigma)|_{\sigma=0,\pi}=\pm \delta X _L(\tau+\sigma)|_{\sigma=0,\pi}$.

Eq. (\ref{nbc1}) is the  usual left and right identification of an open string in a trivial background.  However, this theory  has a non trivial background made of  constant fields $G$ and $B$. These latter are hidden in the definition of the $X_{R;L}$
coordinates. In order to make explicit such a dependence, it is convenient to introduce the
{\em rotated} coordinates $X_{R;L}= O_{R;L} \hat{X}_{R;L}$ satisfying the constraint  $\left.\partial_- {\hat X}_R\right|_{\sigma=0}^{\sigma=\pi}=\pm\left.\partial_+
{\hat X}_L\right|_{\sigma=0}^{\sigma=\pi}$.  One can easily see that  eq. (\ref{nbc0}) is satisfied through the use of these rotated coordinates,  if the invertible matrices $O_{R;L}$ are related by the identity: $O_R^tG^{-1} O_R=O_L^t G^{-1}O_L$. This latter condition, once introduced the matrix $O_R\, O_L^{-1}\equiv {\cal R}^{-t}$, becomes:
\begin{eqnarray}
{\cal R}^t\, G\, {\cal R}=G \label{OD}
\end{eqnarray}
that,  when rewritten in the flat system of coordinates by using the space-time vielbein, is nothing but the definition of  orthogonal group. After having introduced the matrix ${\cal R}$, one can write $X_R={\cal R}^{-t}\,O_L\hat{X}_R$ and $ X_L= O_L\,\hat{X}_L$. ${\cal R}$ acts on the $R$-coordinates  as an  $O(D)$-transformation leaving the action  invariant.  This symmetry  can be fixed by performing the following choice:
\begin{eqnarray}
{\cal R}=E^{-t}\, E\label{gc1}
\end{eqnarray}
where $E=G+B$.
With this choice,  the connection with the standard formulation of the bosonic string in the presence of a magnetic field is  straightforward as one can see in a while.

It is simple to see, with the help of the identity $G\,E^{-t}\,E=E\,E^{-t}\,G$, that the matrix ${\cal R}$ satisfies the condition given in eq. (\ref{OD}).
 The boundary conditions for the  coordinates $X_{L}$ and $X_{R}$ become:
\begin{eqnarray}
\left.\partial_- { X}_R\right|_{\sigma=0}^{\sigma=\pi}=\pm\left.{\cal R}^{-t}\partial_+
{ X}_L\right|_{\sigma=0}^{\sigma=\pi}\Rightarrow E^t\,G^{-1} \left.\partial_- { X}_R\right|_{\sigma=0}^{\sigma=\pi}=\pm\left.E\, G^{-1}\partial_+
{ X}_L\right|_{\sigma=0}^{\sigma=\pi}\label{104}
\end{eqnarray}
which are the standard ones satisfied by an open string in the presence of a Kalb-Ramond field\cite{0709.4149}.
By using  the definition of ${\cal R}$ and eq. (\ref{gc1}),  the matrices $O_{R;L}$ can be written in the form
\begin{eqnarray}
O_L=E^t\, { A}~~;~~O_R=E\, { A}
\end{eqnarray}
being $A\equiv (A)^\mu_{~\nu}$,  at this level, a completely arbitrary matrix. This arbitrarity corresponds to the residual symmetry allowed by gauge choice (\ref{gc1}). On-shell $(\hat{X}_R,\,\hat{X}_L)$ are determined by the equations of motion and the boundary conditions. $(X_R,\,X_L)$, instead, are still  arbitrary because of the ambiguity in the choice of $A$. In order to analyse this extra symmetry, it is interesting to study the transformations induced on such coordinates by changing $A$ and keeping $(\hat{X}_R,\,\hat{X}_L)$ fixed. In detail,  by performing different choices for such  matrices, one can write:
\begin{eqnarray}
X_{R;L}= (G\pm\,B)A_1\hat{X}_{R;L}~~;~~X'_{R;L}= (G\pm\,B)A_2\hat{X}_{R;L}\,\,  .
\end{eqnarray}
The latter equations determine the following transformation both on $X_{R}$ and  $X_{L}$:
\begin{eqnarray}
X_{R;L}=  [(G\pm\,B)A_1A_2^{-1}(G\pm\,B)^{-1}]X'_{R;L} \,\,
\end{eqnarray}
under which the string action has to be invariant,  which happens if:
\begin{eqnarray}
 (A_1A_2^{-1})^t\,{\cal G}_{open} \,A_1A_2^{-1}= {\cal G}_{open}  \,\, . \label{gopen}
\end{eqnarray}
The quantity $ {\cal G}_{open}=E\,G^{-1}\,E^t=E^t\,G^{-1}\,E$ is the so-called {\em open string metric}\cite{0709.4149}. By writing eq. (\ref{gopen}) alternatively for the peculiar cases $(A_1,\,A_2)=(\mathbb{I},\, A_2)$ and $(A_1,\,A_2)=( A_1,\,\mathbb{I})$, one sees that the residual gauge symmetries are the ones that leave the open string metric invariant.

Now that all the ingredients have been introduced, it is straightforward to solve  the equations of motion with the boundary conditions given in eq. (\ref{nbc1}). The solution with the same boundary conditions at $\sigma=0$ and $\sigma=\pi$, i.e.:
\begin{eqnarray}
 \left.\partial_- {\hat X}_R\right|_{\sigma=0}=\pm\left.\partial_+
{\hat X}_L\right|_{\sigma=0}~~;~~\left.\partial_- {\hat X}_R\right|_{\sigma=\pi}=\pm\left.\partial_+
{\hat X}_L\right|_{\sigma=\pi}
 \end{eqnarray}
 can be taken from Ref.\cite{0709.4149}. By writing $X_{R}=E\,\hat{X}_{R}$ and $X_L=E^t\hat{X}_L$ ($A=\mathbb{I}$) one has
\begin{eqnarray}
\hat{X}_R&=& x_R+ \frac{l^2}{\sqrt 2} {\cal G}_{open}^{-1} p(\tau-\sigma) + i \frac{l}{\sqrt 2} \sum_{n\neq 0} \frac{\alpha_n}{n} e^{-in(\tau-\sigma)}\\
\hat{X}_L&=& x_L\pm \frac{l^2}{\sqrt 2} {\cal G}_{open}^{-1} p(\tau+\sigma) \pm i \frac{l}{\sqrt 2} \sum_{n\neq 0} \frac{\alpha_n}{n} e^{-in(\tau+\sigma)} \, .
\end{eqnarray}
For mixed boundary conditions
\begin{eqnarray}
\left.\partial_- {\hat X}_R\right|_{\sigma=0}=\pm\left.\partial_+
{\hat X}_L\right|_{\sigma=0}~~;~~\left.\partial_- {\hat X}_R\right|_{\sigma=\pi}=\mp\left.\partial_+ {\hat X}_L\right|_{\sigma=\pi}
\end{eqnarray}
one instead obtains:
\begin{eqnarray}
\hat{X}_R&=& x+ i \frac{l}{\sqrt 2} \sum_{r\in \mathbb{Z}+\frac{1}{2} } \frac{\alpha_r}{r} e^{-ir(\tau-\sigma)} \,\, ,\\
\hat{X}_L&=& x \pm i \frac{l}{\sqrt 2} \sum_{r\in \mathbb{Z}+\frac{1}{2}} \frac{\alpha_r}{r} e^{-ir(\tau+\sigma)} \,\, .
\end{eqnarray}
The mode expansion of the starting $(X,\tilde{X})$ coordinates are given by:
\begin{eqnarray}
X=\frac{1}{\sqrt{2}}\left(G^{-1} E \hat{X}_R+ G^{-1} E^t\hat{X}_L\right) ~;~\tilde{X}=\frac{1}{\sqrt{2}}\left(EG^{-1} E \hat{X}_R-E^t G^{-1} E^t\hat{X}_L\right)  .\label{xxtilde}
\end{eqnarray}

The expression of $X$ given in the first identity of eq. (\ref{xxtilde}) coincides with the standard open string expansion in the presence of a Kalb-Ramond field \cite{0709.4149}.
The second identity in the same equation is its dual expression. In order to have a more intuitive picture of what ``dual field'' means in this context,  it is enlightening to consider the case $B=0$ as an example. When $B=0$ then $(X_R,\,{X}_L)=(G\,\hat{X}_R,\,G\,\hat{X}_L)$, and   the mode expansions  of the  $X$ and $\tilde{X}$-fields simplify being equal to:
\begin{eqnarray}
X&=&x+ \frac{l^2}{4}G^{-1}p[\tau-\sigma\pm(\tau+\sigma)]+i\frac{l}{2}\sum_{n\neq0}e^{-in\tau} \frac{\alpha_n}{n}\left( e^{in\sigma}\pm e^{-in\sigma}\right)\label{fex}\\
\tilde{X}&=&\tilde{x} +\frac{l^2}{4}p[\tau-\sigma \mp(\tau+\sigma)] +i\frac{l}{2}\sum_{n\neq0}e^{-in\tau} \frac{\alpha_n}{n}\left( e^{in\sigma}\mp e^{-in\sigma}\right)\label{fex1}
\end{eqnarray}
with $x=(x_R+x_L)/\sqrt{2}$ and $\tilde{x}=G\, (x_R-\,x_L)/\sqrt{2}$. The expression given in eq. (\ref{fex}), taken with the upper sign, in the following denoted by $X^{(+)}$,  is the usual mode expansion of the string coordinates having  NN-boundary conditions while  the expression with the lower sign, i.e. $X^{(-)}$,  corresponds to open strings
 with DD-boundary conditions. For the $\tilde{X}$, given in eq. (\ref{fex1}), this correspondence  is inverted.  In particular, by  denoting again by  $\tilde{X}^{(\pm)}$  the two expressions associated respectively with the upper and lower choice of the signs in eq. (\ref{fex}), one finds the suggestive identity:
\begin{eqnarray}
 X^{(+)}=G^{-1} \tilde{X}^{(-)}~~;~~ X^{(-)}=G^{-1} \tilde{X}^{(+)} \,\, .
 \end{eqnarray}
These are  the expected relations for T-dual coordinates in absence of the Kalb-Ramond $B$-field.

The quantization of this system is not straightforward because now the $X_R$ and $X_L$ fields are not any more  independent, for this reason its study is postponed in a forthcoming publication.

\vspace{.2cm}

\begin{center}
\bf{Acknowledgments}
\end{center}
The authors are deeply indebted to Olaf Hohm, Magdalena Larfors and Dmitri Sorokin for fruitful discussions on different issues faced in this paper.

G. G. is grateful to Bill Stoeger S. J. for encouraging him to work in this subject.

F. P. would like to thank Barton Zwiebach for stimulating conversations held at the MIT Center for Theoretical Physics (INFN-MIT {\em Bruno Rossi} exhange program). He also would like to thank the SGGP for hospitality and support during the 11th Simons Workshop on Mathematics and Physics.

R. M. and F. P. would like to thank the Galileo Galilei Institute in Florence for hospitality during the Workshop on {\em Geometry  of String and Fields}. Furthermore, both of them are grateful to the Dipartimento di Fisica of Federico II University in Naples, for hospitality and support.

\appendix

\section{Notations and useful identities}
It is useful to summarize the notation adopted for the indices. Two-dimensional flat indices are denoted by $a, b,\dots$; the corresponding curved ones are denoted by the Greek letters $\alpha,\beta,\dots$ $\,$.
The indices used for labelling the $2D$ compact dimensions are $i, j, \dots$, while the ones adopted for the $D$ compact directions are $\mu,\nu,\dots$ $\,$.

The Fourier expansion of the Dirac delta function is:
\begin{eqnarray}
\sum_{n\in\mathbb{Z}} e^{2 i n\sigma} = 2 \pi \delta (2 \sigma) = \pi \delta (\sigma) \,\, , \qquad \sigma\in[0, \,\pi] \, .\label{ds}
\end{eqnarray}
It is connected with the Heaviside $\theta$-function by the identity:
\begin{eqnarray}
\epsilon(\sigma) \equiv \frac{1}{2}\left[ \theta(\sigma)-\theta(-\sigma)\right]=\frac{1}{2}\int_{-\sigma}^\sigma \, dt \,\, \delta(t)
= \frac{2\sigma}{2\pi} -\frac{i}{2\pi} \sum_{n\neq 0} \frac{1}{n} e^{2 i n \sigma} \, . \label{es}
\end{eqnarray}
The above equation implies that $\partial_\sigma\epsilon ( \sigma ) = \delta ( \sigma)$.

The following notation for the Poisson brackets at equal $\tau$, introduced for the first time in eq. (\ref{pb}), has been used:
$$
\left\{ P_{R;L}(\sigma),X_{R;L}^{t}(\sigma') \right\}_{PB} \equiv
$$
\begin{eqnarray}
\left( \begin{array}{cccc}
\left\{ P_{R;L}^{1}(\sigma),\,X_{R;L\,1}(\,\sigma') \right\} & \left\{ P_{R;L}^{1}(\sigma),\,X_{R;L\,2}(\,\sigma') \right\} & \dots & \left\{P_{R;L}^{1}(\,\sigma),\,X_{R;L \, D}(\,\sigma')\right\} \\
\left\{ P_{R;L}^{2}(\sigma),\,X_{R;L\,1}(\,\sigma') \right\} & \left\{ P_{R;L}^{2}(\sigma),\,X_{R;L\,2}(\,\sigma') \right\} & \dots & \left\{P_{R;L}^{2}(\,\sigma),\,X_{R;L \, D}(\,\sigma')\right\} \\
\vdots & \vdots & \ddots & \vdots \\
\left\{ P_{R;L}^{D}(\,\sigma),\,X_{R;L\,1}(\,\sigma')\right\} & \left\{P_{R;L}^{D}(\,\sigma),\,X_{R;L\,2}(\,\sigma')\right\} & \dots & \left\{P_{R;L}^{D}(\,\sigma),\,X_{R;L\,D}(\,\sigma')\right\}  \nonumber
\end{array}
\right)  .
\end{eqnarray}

The Dirac brackets for the right sector are defined as:
\begin{eqnarray}
\left\{\cdot \, , \cdot \right\}_{DB}&=&\left\{\cdot \, ,\cdot \right\}_{PB}-\int d\sigma \, d\sigma' \left\{\cdot\, , \Psi_R^t \right\}_{PB}\left[ \left\{\Psi_R, \Psi_R^{'t}\right\}_{PB}\right]^{-1} \left\{\Psi_R^{'},\cdot\right\}_{PB}\nonumber \\
&=&\left\{\cdot \, ,\cdot \right\}_{PB}-\int d\sigma \, d\sigma' \left\{\cdot \, , \Psi_R^t\right\}_{PB}\, \left[-\frac{G}{T} \epsilon(\sigma-\sigma')\right] \, \left\{\Psi_R^{'}, \cdot\right\}_{PB} \, ,
\end{eqnarray}
(where $\Psi_R \equiv \Psi_R (\tau , \sigma)$ and $\Psi'_R \equiv \Psi_R (\tau, \sigma')$) with a similar expression for the left sector. In the latter equation the derivative of the Dirac $\delta$-function, which was in $\left\{\Psi_R, \Psi_R^{'t}\right\}_{PB}$, has been subsituted by the step function $\epsilon (\sigma- \sigma')$. This is possible due to the following integral identity:
\begin{eqnarray}
\int d\tilde{\sigma}[\partial_\sigma \delta(\sigma-\tilde{\sigma})]\epsilon(\tilde{\sigma}-\sigma')=\partial_\sigma\epsilon({\sigma}-\sigma') =\delta(\sigma-\sigma') \, .
\end{eqnarray}
It shows that the $\epsilon$-function is the ``inverse'' of $\partial_\sigma \delta ( \sigma)$ .

\section{Equations of motion, symmetries and quantization} \label{B}

In the Tseytlin double sigma model, the equations of motion for the fields $\chi^i$ are obtained from the variation of the action given in eq.(\ref{act}):
\begin{eqnarray}
\delta S &=&- \int d^2\xi\, \partial_\alpha\left[e \,e_1^{~\alpha}\, \delta \chi^t\left( C\nabla_0\chi+M\nabla_1 \chi \right)\right] -\frac{1}{2} \int d^2\xi\, \partial_\alpha\left(\epsilon^{\alpha\beta}\delta \chi^t C\,\partial_\beta\chi\right)\nonumber\\
& &+ \int d^2\xi \, \delta \chi^t \,\left\{\partial_\alpha \left[e \,e_1^{~\alpha}\left( C\nabla_0\chi+M\nabla_1\chi\right)\right] \right\}\, .\label{vact}
\end{eqnarray}
Here, $\nabla_a$ is a linear combination of covariant derivatives ($ \nabla_a \equiv e_a^{~\alpha}\nabla_\alpha$), and $\nabla_\alpha$, acting on a world-sheet scalar, can be equivalently thought as the two-dimensional covariant derivative or the usual partial one.

The first two integrals in eq. (\ref{vact}) give the following boundary terms:
\begin{eqnarray}
\delta S_{boundary} &=&- \int d\tau \left. \left\{ \delta \chi^t \left[\,e\, e_1^{~1}\left( C\nabla_0\chi+M\nabla_1 \chi \right)- \frac{1}{2} C\, \partial_0\chi\right] \right\} \right|_{\sigma=0}^{\sigma=\pi} \, ,
\end{eqnarray}
while the last one in eq. (\ref{vact}) gives the equation of motion:
\begin{eqnarray}
\partial_\alpha \left[e \, e_1^{~\alpha} \left( C\nabla_0\chi+M\nabla_1\chi\right)\right]=0 \, .
\end{eqnarray}

If one performs the following shift in the fields
\begin{eqnarray}
\chi(\xi^\alpha)\rightarrow \chi'(\xi^\alpha)\equiv\chi(\xi^\alpha)+f(\xi^\alpha)~,~~~ \mbox{with}~~ \nabla_1f=0\label{shift}
\end{eqnarray}
the action (\ref{act}) acquires a boundary term
\begin{eqnarray}
S\rightarrow S'\equiv S-\frac{1}{2}\int d^2 \xi\epsilon^{\alpha\beta}\partial_\alpha f^t\,C \partial_\beta \chi
\end{eqnarray}
In the flat gauge, where $e^a_{~\alpha}=\delta^a_\alpha$, one has: $f\equiv f(\tau)$.

The equations of motion obtained from this modified action are unchanged while the boundary terms get modified:
\begin{eqnarray}
\delta S_{boudary}\rightarrow\delta S'_{boundary}\equiv \delta S_{boundary}-\left.\frac{1}{2}\int d\tau \left[ \delta \chi^t \, C \, \partial_0 f \right] \right|_{\sigma=0}^{\sigma=\pi} \, .\label{bomo}
\end{eqnarray}
In the double closed string theory,
by assuming that the function $f$ satisfy the same periodic identification $f(\tau,\,\sigma+\pi) \equiv f(\tau,\,\sigma)$ as the function $\chi$, the last term in eq. (\ref{bomo}) vanishes, so proving the invariance of both the equations of motion and the boundary terms under the shift symmetry showed in eq. (\ref{shift}). The fields $\chi'$, $\chi$ and, for consistency, also the function $f$, satisfy the same boundary conditions. This remark justifies the periodic identification imposed on the vector function $f$ and allows to cancel the boundary term also when open string like boundary conditions are imposed.

The components of the tensor $t_a^{~b}$ can be easily read from the action (\ref{act}):
\begin{eqnarray}
t_{0}^{\,\,0}  = - M_{ij} \nabla_{1} \chi^{i} \nabla_{1} \chi^{j} ~~~~~~~~~~~~~~~~~~~~~~&;&~~ t_{1}^{\,\,1}  =   M_{ij} \nabla_{1} \chi^{i} \nabla_{1} \chi^{j}\nonumber \\
t_{0}^{\,\, 1} =   C_{ij} \nabla_{0} \chi^{i} \nabla_{0} \chi^{j}  +2 M_{ij} \nabla_{0} \chi^{i} \nabla_{1} \chi^{j}~~&;&~~ t_{1}^{\,\, 0}  =   C_{ij} \nabla_{1} \chi^{i} \nabla_{1} \chi^{j}
\end{eqnarray}
In the light-cone gauge they become
\begin{eqnarray}
&&t_{++}= \frac{1}{2}t_{00} +\frac{1}{4}\left(t_{01}+t_{10} \right)~~;~~t_{--}= \frac{1}{2}t_{00} -\frac{1}{4}\left(t_{01}+t_{10} \right)\nonumber\\
&&t_{+-}=-t_{-+}=-\frac{1}{4}(t_{01}-t_{10})=-\frac{1}{4} \epsilon^{ab}t_{ab} \,\, .
\end{eqnarray}

It can be useful to show that the mode expansions given in eqs.  (\ref{seqx}) and (\ref{seqx1}), with the parentheses defined in eq. (\ref{dirbra}), satisfy the Dirac brackets written in eqs. (\ref{RLDB}) or (\ref{DCB}):
\begin{eqnarray}
\left\{{X_{R;L}}_\mu (\tau,\,\sigma),\,{X_{R;L}}_\nu (\tau,\,\sigma')\right\}_{DB} &=&\mp 2\,l^2\,G_{\mu \nu}\left[(\sigma-\sigma')
-\frac{i}{2} \sum_{n\neq 0} \frac{1}{n}e^{2in(\sigma-\sigma')}\right]\nonumber\\
&=&\mp 2\pi\,l^2\, G_{\mu \nu} \epsilon(\sigma-\sigma')\nonumber\\
\left\{P_{R;L}^\mu (\tau,\,\sigma),\,{X_{R;L}}_\nu (\tau,\,\sigma')\right\}_{DB}&=&\delta^\mu_\nu T\,l^2\sum_{n\in\mathbb{Z}}  e^{2in(\sigma-\sigma')}=\frac{1}{2}\delta(\sigma-\sigma')\nonumber\\
\left\{P_{R;L}^\mu (\tau,\,\sigma),\,P_{R:L}^\nu (\tau,\,\sigma')\right\}_{DB}&=&\pm T^2\,l^2 G^{\mu\nu}\, i \sum_{n\neq 0} n\,e^{2in(\sigma-\sigma')}\nonumber\\
&=&\pm G^{\mu \nu} \frac{T}{4} \delta'(\sigma-\sigma')
\end{eqnarray}
being, for the Tseytin's action given in sect. \ref{tseytlin}, $P_{R;L}=\mp\frac{T}{2}G^{-1}\,\partial_1X_{R;L}$ while for the Hull's one given  in sect. \ref{hull}, the expression of the conjugate momenta is $P_{R;L}=\frac{T}{2}G^{-1}\,\partial_0X_{R;L}$. In both cases one can write:
\begin{eqnarray}
P_R= T \,l\, G^{-1}\sum_{n\in \mathbb{Z}} \alpha_ne^{-2in(\tau-\sigma)}~~;~~
P_L= T \,l\, G^{-1}\sum_{n\in \mathbb{Z}} \tilde{\alpha}_ne^{-2in(\tau+\sigma)}
\end{eqnarray}
with $\alpha_0= l\,p_R$, $\tilde{\alpha}_0=lp_L$ and $T=1 / (2\pi\,l^2)$.

The action of the $O(D,D)$ group on the coordinates $\Phi$ introduced in sect. \ref{tseytlin}, is better understood through the target space vielbein, defined by $G_{\mu\nu}=E_\mu^{~a}\delta_{ab}E_\nu^{~b}$.
The $O(D,D)$ metric is now written as:
\begin{eqnarray}
{\cal C}=\left(\begin{array}{cc}
                  E_\mu^{~a}&0\\
                  0&  E_\mu^{~a}\end{array}\right) \left( \begin{array}{cc}
                                                             \delta_{ab}&0\\
                                                             0&-\delta_{ab}\end{array} \right) \left(\begin{array}{cc}
                  E_\nu^{~b}&0\\
                  0&  E_\nu^{~b}\end{array}\right)\equiv E\,C\,E^t \, ,
\end{eqnarray}
being $C$ the matrix defined in eq. (\ref{24}). A matrix ${\cal R}$, belonging to the  non-compact orthogonal group, acts on the coordinates $\tilde{\Phi}=E^{-1}\Phi$ as ${\tilde \Phi}'={\cal R}\,\tilde{\Phi}$ and leaves $C$ invariant, i.e.  ${\cal R}^{-1}\, C\,{\cal R}^{-t}=C$. The matrix ${\cal G}$ in the flat system of coordinates becomes, instead, the identity matrix:
\begin{eqnarray}
{\cal G}=\left(\begin{array}{cc}
                  E_\mu^{~a}&0\\
                  0&  E_\mu^{~a}\end{array}\right) \left( \begin{array}{cc}
                                                             \delta_{ab}&0\\
                                                             0&\delta_{ab}\end{array} \right) \left(\begin{array}{cc}
                  E_\nu^{~b}&0\\
                  0&  E_\nu^{~b}\end{array}\right) \, \, .
\end{eqnarray}
 It is not invariant under an $O(D,D)$ transformation. In other words, the matrix ${\cal G}$ is no longer of the form given in eq. (\ref{trgo}) after the action of an element of such non-compact group.

The Dirac brackets can be expressed in a more simplified notation by introducing the vector  $\Phi=(X_R,\,X_L)$ and ${\cal P}=(P_R,\,P_L)$:
\begin{eqnarray}
\left\{ \Phi(\tau,\sigma),\,\Phi^t(\tau,\,\sigma')\right\} &=& \frac{1}{T} \, {\cal C}\,  \epsilon(\sigma-\sigma') \nonumber \\
\left\{ {\cal P} (\tau,\sigma),\,\Phi^t(\tau,\,\sigma')\right\} &=& \, \frac{1}{2}\,\mathbb{I} \, \delta(\sigma-\sigma') \label{comm} \\
\left\{ {\cal P} (\tau,\sigma),\,{\cal P}^t(\tau,\,\sigma')\right\} &=& \frac{T}{4} \,{\cal C}^{-1} \, \delta'(\sigma-\sigma') \, . \nonumber
\end{eqnarray}
It is also useful to rewrite these brackets in terms of the original variables ${\cal X}$ and ${\cal P}$. This is obtained by writing the conjugate momenta in the Tseytlin and Hull theories in the chiral basis and transforming them in the original basis where the coordinates are $X$ and $\tilde{X}$. In detail, the conjugate momenta are, respectively:
\begin{eqnarray}
P=\frac{T}{2} {\cal C}^{-1} \,\partial_1\Phi ~~;~~P=-\frac{T}{2} {\cal G}^{-1}\,\partial_0 \Phi \, \, .
\end{eqnarray}
By using the identity $\Phi={\cal T}\,{\cal X}$, where ${\cal T}$ is the matrix whose inverse is defined in eq. (\ref{calT}), and the identities written in eq. (\ref{trgo}), one has in both the theories:
\begin{eqnarray}
{\cal P}= {\cal T}^t\,P= \frac{T}{2}\, \Omega\, \partial_1{\cal X}~~;~~{\cal P}= {\cal T}^t\,P=\frac{T}{2} \,M\,\partial_0{\cal X} \,\, .
\end{eqnarray}
The previous identities allow to write:
\begin{eqnarray}
&&{\cal T}^{-1}\,\left\{ \Phi(\tau,\sigma),\,\Phi^t(\tau,\,\sigma')\right\}{\cal T}^{-t} =  \left\{ {\cal X}(\tau,\sigma),\,{\cal X}^t(\tau,\,\sigma')\right\} \nonumber \\
&& =   \frac{\epsilon(\sigma-\sigma')}{T} {\cal T}^{-1}\,{\cal C}\,{\cal T}^{-t}= \frac{\epsilon(\sigma-\sigma')}{T}\Omega^{-1}
\end{eqnarray}
which is the first Dirac brackets written in eq. (\ref{RLDB}). The other parentheses given in  the same equation are similarly obtained.

\section{Open strings and the $O(D)$ symmetry}

In this Appendix the role of the $O(D)$ symmetry discussed in sect. \ref{openstrings} is examined from a different point of view. In particular it is shown that the solution of the equations of motion with boundary conditions written in eq. (\ref{nbc1}) and the ones obtained by imposing the left and right identification written in eq. (\ref{104}) are related by the $O(D)$ matrix ${\cal R}^t=E^t\,E^{-1}$.

The boundary conditions shown in eq. (\ref{nbc1}) do not exhibit any dependence on the $B$-field. As a consequence, the solutions of the duality equations with such boundary conditions are:
\begin{eqnarray}
{X}'_R&=& q_R+ \frac{l^2}{\sqrt 2} \,
q(\tau-\sigma) + i \frac{l}{\sqrt 2} \sum_{n\neq 0} \frac{a_n}{n}
e^{-in(\tau-\sigma)} \nonumber \\
\label{los1} \\
{X}'_L&=& q_L\pm \frac{l^2}{\sqrt 2}
q(\tau+\sigma) \pm i \frac{l}{\sqrt 2} \sum_{n\neq 0} \frac{a_n}{n}
e^{-in(\tau+\sigma)}\, . \nonumber
\end{eqnarray}
On the other hand, the solution of the duality equations with boundary conditions written in eq. (\ref{104}) are ($A=\mathbb{I}$):
\begin{eqnarray}
{X}_R&=& E x_R+ \frac{l^2}{\sqrt 2} E\,{\cal G}_{open}^{-1}
\,p(\tau-\sigma) + i \frac{l}{\sqrt 2} \sum_{n\neq 0}
\frac{E\,\alpha_n}{n} e^{-in(\tau-\sigma)}\nonumber \\
\label{los2} \\
{X}_L&=& E^t\,x_L\pm \frac{l^2}{\sqrt 2}E^t\, {\cal G}_{open}^{-1}
\,p(\tau+\sigma) \pm i \frac{l}{\sqrt 2} \sum_{n\neq 0}
\frac{E^t\,\alpha_n}{n} e^{-in(\tau+\sigma)} \, . \nonumber
\end{eqnarray}
The solutions given in eqs. (\ref{los1}) and (\ref{los2}) have to be related because  eq. (\ref{nbc1}) is obtained from eq. (\ref{104})  by taking ${\cal R}=\mathbb{I}$.

The simplest connection between the two solutions is obtained by identifying $X_R=X'_R$. This latter condition determines the following relations among the Fourier modes, $ q_R=E\,x_R$, $q=
E\,{\cal G}^{-1}\,p_R$ end $a_n=E\,\alpha_n$. These latter relations
once used in the expression of $X_L$ gives:
\begin{eqnarray}
  {X}_L=E^t\, E^{-1}X'_L \Longrightarrow  X_L={\cal R}^t {X}'_L
\end{eqnarray}
where it has been set $ q_L=E\,x_L$ by analogy with the right coordinates.
 From the previous expression one sees that the fields $L$ are related by the matrix ${\cal R}^t$ and this property is in agreement with the general symmetry arguments introduced in sect.~\ref{openstrings}.

\end{document}